\documentclass[apj]{emulateapj}
\usepackage{amsmath}
\usepackage{graphicx}
\usepackage{enumitem}
\usepackage{natbib}
\usepackage{longtable}
\usepackage{url}

\begin{document}

\title{Flexible and Scalable Methods for Quantifying Stochastic Variability in the Era of Massive Time-Domain Astronomical Data Sets}
\shorttitle{Quantifying Stochastic Variability}
\shortauthors{Kelly et al.}

\author{Brandon C. Kelly\altaffilmark{1}, Andrew C. Becker\altaffilmark{2}, Malgosia Sobolewska\altaffilmark{3}, Aneta Siemiginowska\altaffilmark{4}, Phil Uttley\altaffilmark{5}}

\altaffiltext{1}{Department of Physics, Broida Hall, University of California, Santa Barbara, CA 93106-9530}
\altaffiltext{2}{Department of Astronomy, University of Washington, P.O. Box 351580, Seattle, WA 98195-1580}
\altaffiltext{3}{Nicolaus Copernicus Astronomical Center, Bartycka 18, 00-716, Warsaw, Poland}
\altaffiltext{4}{Harvard-Smithsonian Center for Astrophysics, 60 Garden St, Cambridge, MA 02138}
\altaffiltext{5}{Astronmical Institute `Anton Pannekoek', University of Amsterdam, Postbus 94249, 1090 GE Amsterdam, the Netherlands}

\begin{abstract}
We present the use of continuous-time autoregressive moving average (CARMA) models as a method for estimating the variability features of a light curve, and in particular its power spectral density (PSD). CARMA models fully account for irregular sampling and measurement errors, making them valuable for quantifying variability, forecasting and interpolating light curves, and for variability-based classification. We show that the PSD of a CARMA model can be expressed as a sum of Lorentzian functions, which makes them extremely flexible and able to model a broad range of PSDs. We present the likelihood function for light curves sampled from CARMA processes, placing them on a statistically rigorous foundation, and we present a Bayesian method to infer the probability distribution of the PSD given the measured lightcurve. Because calculation of the likelihood function scales linearly with the number of data points, CARMA modeling scales to current and future massive time-domain data sets. We conclude by applying our CARMA modeling approach to light curves for an X-ray binary, two AGN, a long-period variable star, and an RR-Lyrae star, in order to illustrate their use, applicability, and interpretation.
\end{abstract}


\section{Introduction}
\label{s-intro}

Current and future time domain optical surveys, such as the SDSS Stripe 82 Supernova Survey \citep{Frieman2008a}, Palomar Transient Factory \citep[PTF,][]{Law2009a}, the Catalina Real-Time Transient Survey \citep[CRTS,][]{Drake2009a}, Pan-STARRS \citep{Kaiser2002a}, and the Large Synoptic Survey Telescope \citep[LSST,][]{Ivezic2008a}, are providing and will provide an unprecedented flood of variability data. Such data sets will number in the millions to hundreds of billions of photometric data points, providing systematic multi-wavelength variability studies for thousands to eventually billions of objects. For example, LSST will have $\sim 30$ trillion total measurements on $\sim 40$ billion objects. For many classes of astronomical sources these will be the first systematic multi-passband variability studies including large numbers of objects with well-sampled multiwavelength light curves. Moreover, these rich data sets will also enable the inclusion of multi-passband variability information for distinguishing different classes of objects. Data sets generated by these surveys will present many exciting opportunities, providing astrophysical insight for known classes of objects, as well as the discovery of unknown variability classes, new subclasses of known variable classes, and anomalous outliers.

The central data analysis problem for extracting science from these time-domain data sets is how to quantitatively characterize the variability. For periodic signals characterizing the variability is relatively straightforward, with the period being the obvious and most important feature to use. For transient phenomenon, such as supernovae or black hole tidal disruptions, the light curve is often characterized by fitting a parameterized deterministic model to the data, either statistical (e.g., splines) or astrophysical. For quasi-periodic and stochastic light curves the variability is often characterized through the power spectral density (PSD). The PSD is the variability amplitude per frequency, so it describes the variability power contained within a frequency interval. A similar measure that is sometimes used is the structure function, which describes the variability amplitude as a function of time scale. The variability of quasi-periodic and stochastic light curves may then be characterized by summarizing the power spectrum through a parametric form, with power-laws and sums of Lorentzian functions being common choices for AGN and X-ray binaries, respectively.

Irregular sampling or sequences of regular sampling separated by gaps are often the source of the most problematic aspects of measuring variability features from a lightcurve. Unfortunately, all ground-based astronomical data and many space-based data are subject to sampling which is not strictly uniform. For periodic signals, methods have been developed to estimate periods from irregularly sampled light curves, and assess their statistical significance \citep[e.g.,][]{Scargle1982a,Horne1986a,Reimann1994a}; significance tests have also been developed for periodic signals against red noise for regularly sampled light curves \citep{Vaughan2010a}. For deterministic models, fitting the light curve is a traditional regression problem so the sampling pattern does not bias the results. 

Deriving the PSD features and their uncertainties from an irregularly sampled light curve is considerably more challenging for stochastic light curves. For a regularly sampled light curve the traditional way of estimating the PSD is through the Discrete Fourier Transform of the light curve, the modulus-squared of which is called the periodogram. The periodogram suffers from biases due to the fact that it is calculated from a time series that is a sample from a continuous-time stochastic process. As a result, the sampling pattern of the light curve distorts the periodogram relative to the true PSD that generated the light curve \citep[e.g.,][]{Uttley2002a, Vaughan2003a}. These issues similarly cause the empirical structure function to be a distorted estimate of the true structure function \citep{Emmanoulopoulos2010a}. This is a concern because if one does not correct for this distortion, differences in variability properties caused by vagaries of the sampling pattern could mistakenly be interpreted to have an astrophysical origin. Moreover, this distortion is also a problem for classification algorithms that utilize variability information, as objects may spuriously be interpreted to belong to different variability classes simply because they have different sampling patterns. Clearly this issue must be dealt with in order to take advantage of the state of the art time domain data sets. 

There are two primary approaches used in astronomical studies to account for irregular sampling in stochastic light curves. The first approach is to use Monte Carlo simulations to forward model the periodogram as a function of a model power spectrum \citep{Done1992a,Uttley2002a,Emmanoulopoulos2013a}. The approach proceeds by first simulating a large number of light curves from an assumed power spectrum, sampling them with the same pattern as the measured light curve, computing the periodogram for each down sampled simulated light curve, and averaging the simulated periodograms to create a `model' periodogram. The best-fit power spectrum is found by minimizing the $\chi^2$ between the model periodogram and the measured periodogram. Moreover, confidence intervals may be estimated from the Monte Carlo samples. This approach is extremely flexible, but can be computationally expensive. This is especially true if there are intervals of fine sampling separated by intervals of sparse sampling, as this requires either generating a very dense light curve at the finest sampling rate, or splitting the light curve into segments and computing their periodograms separately. Unfortunately, because of the computational cost, it is difficult to see how this approach can be applied to massive time domain data sets. Moreover, fitting complex multi-component for even a single data set can become computationally expensive in the Monte Carlo forward fitting approach.

The second approach to accounting for irregular sampling in stochastic light curves is to fit the light curve in the time domain. This is almost always done by assuming the light curve is a realization of a Gaussian process \citep[e.g.,][]{Rybicki1992a,Kelly2009a,Kelly2011a,Kelly2013b,Miller2010a,Kozowski2010a,Brewer2011a}. In this case, the likelihood function is a multivariate normal distribution with unknown mean and covariance matrix. The covariance matrix is parameterized by the autocovariance function, which forms a Fourier transform pair with the power spectrum. In this time-domain approach, the autocovariance function is fit through maximum-likelihood or Bayesian approaches, and the power spectrum is calculated from the inferred autocovariance function. The irregular sampling and measurement errors are automatically accounted for by the likelihood function. This approach is statistically powerful because of its reliance on the likelihood function, but in general is also computationally expensive. In order to calculate the multivariate Gaussian likelihood function it is necessary to invert the $n \times n$ covariance matrix of the light curve, where $n$ is the number of data points in the light curve. In general matrix inversion scales as $O(n^3)$, which can be prohibitive for large samples of light curves that are sampled with even moderate density.

There are special classes of Gaussian processes for which the computational complexity only scales linearly with the length of the light curve. For linear processes which have a `state-space' representation \citep[discussed further in \S~\ref{s-statespace}, see also][]{Vio1992a} the likelihood function can be computed using a computationally efficient (scaling as $O(n)$) algorithm known as the Kalman Filter. Gaussian processes with an exponential autocorrelation fall into this class of stochastic models that have fast algorithms for computing their likelihood function \citep{Rybicki1995a,Kozowski2010a}. This particular process is known as both a first order continuous-time autoregressive process (CAR(1)) or an Ornstein-Uhlenbeck process\footnote{There is effectively no difference between an Ornstein-Uhlenbeck process and a first-order continuous time autoregressive process, although the terminology Ornstein-Uhlenbeck process is more common in the physics literature.}, and was introduced by \cite{Kelly2009a} as a model for quasar optical light curves. Subsequent work has confirmed that it provides a good description of quasar optical light curves on time scales of days to years at the level of data quality of the OGLE and Stripe 82 surveys \citep{Kozowski2010a,MacLeod2010a,Zu2013a,Andrae2013a}. However, recent studies have found evidence for deviations from the CAR(1) model for optical light curves of AGN \citep{Mushotzky2011a,Graham2014a}. In addition, successful AGN selection techniques have been developed based on the CAR(1) parameters \citep{Butler2011a,MacLeod2011a,Ruan2012a,Choi2013a}.

Despite its recent success, the CAR(1) model is very simple, as it assumes a PSD that is a Lorentzian centered at zero. Thus, there are only two free parameters in the CAR(1) model: the bend frequency of the PSD (i.e., the width of the Lorentzian) and the normalization. This makes the model inflexible, limiting its broader use. In this paper we overcome the inflexiblity of the CAR(1) model and present the general class of continuous time autoregressive moving average (CARMA) models. CARMA models are generated by adding higher order derivatives to the stochastic differential equation that defines the CAR(1) processes. The special case of a CAR($p$) process is discussed by \citet{Koen2005a}, who applied it to the light curves for two variable stars. The PSD of a CARMA process is a sum of Lorentzian functions, with the free parameters being the centroid, widths, and normalizations of the Lorentzian functions. This provides a significant amount of flexibility in modeling the PSD, making CARMA modeling applicable to many classes of astronomical variables. Moreover, CARMA models have a state space representation, enabling the use of the Kalman Filter for calculating their likelihood function. Because of this, the computational complexity of calculating the likelihood function for CARMA models still scales linearly with the number of data points in the light curve, making them scalable to massive time domain data sets. 

The format of this paper is as follows. In \S~\ref{s-carma_def} we begin by defining the CARMA process via a stochastic differential equation, present the PSD of a CARMA process, and present the autocovariance function of a CARMA process. Then, in \S~\ref{s-statespace} we express the CARMA process using a continuous-time state space representation and use this representation to derive the solution to the stochastic differential equation defining the process. This solution forms the basis for the statistical properties of a CARMA process sampled at a set of observational times, which are necessary for fitting the CARMA process parameters to a measured time series. In \S~\ref{s-inference} we present the likelihood function for a sampled CARMA process, and in \S~\ref{s-likelihood} present an efficient algorithm for computing the likelihood function based on the Kalman filter. In order to derive the Kalman filter for a CARMA process it is necessary to obtain a discrete-time state space representation of the sampled process, and we begin \S~\ref{s-likelihood} by using the results of \S~\ref{s-statespace} to present this discrete time representation. We extend these results in \S~\ref{s-interpolation} and derive an algorithm for efficiently performing interpolation and extrapolation from a measured time series assuming a CARMA model. In \S~\ref{s-bayes} we describe Bayesian inference for the CARMA model, including our adopted prior distribution, in \S~\ref{s-assess_fit} we describe how to assess the quality of fit based on the CARMA model, in \S~\ref{s-order} we discuss how to choose the order of the CARMA model, and in \S~\ref{s-computation} we discuss computational aspects related to fitting the CARMA model to a measured time series. In \S~\ref{s-simulations} we illustrate statistical inference under a CARMA model on two simulated lightcurves, and in \S~\ref{s-applications} we apply the CARMA model to astronomical lightcurves from an X-ray binary, AGN, and periodic variable stars. In \S~\ref{s-discussion} we discuss our results, and provide directions for future development.

\section{Continuous-Time Autoregressive Moving Average Models}
\label{s-carma}

In this section we introduce the important mathematical properties of the CARMA(p,q) process, including its definition, PSD, autocovariance function, and solution. Further details may be found in, e.g., \citet{Jones1981a}, \citet{Jones1990a}, \citet{Belcher1994a}, \citet{roux2002some}, and \citet{Koen2005a}.

\subsection{Definition, Power Spectrum, and Autocovariance Function}
\label{s-carma_def}

A zero-mean CARMA(p,q) process $y(t)$ is defined to be the solution to the stochastic differential equation
\begin{multline}
	\frac{d^p y(t)}{dt^p} + \alpha_{p-1} \frac{d^{p-1} y(t)}{dt^{p-1}} + \ldots + \alpha_0 y(t) = \\
	 \beta_q \frac{d^q \epsilon(t)}{dt^q} + \beta_{q-1} \frac{d^{q-1} \epsilon(t)}{dt^{q-1}} + \ldots + \epsilon(t).
	\label{eq-carma}
\end{multline}
Here, $\epsilon(t)$ is a continuous time white noise process with zero mean and variance $\sigma^2$. In addition, we define $\alpha_p = 1$ and $\beta_0$ = 1. The parameters $\alpha_0,\ldots, \alpha_{p-1}$ are the autoregressive coefficients, and the parameters $\beta_1, \ldots, \beta_q$ are the moving average coefficients. For the process to be stationary, it is necessary that $q < p$ and that the roots $r_1,\ldots,r_p$ of the autoregressive polynomial
\begin{equation}
	A(z) = \sum_{k=0}^p \alpha_k z^k
	\label{eq-characteristic}
\end{equation}
have negative real parts.  In addition, the process has `minimum phase' when the roots of the moving average polynomial have non-positive real parts. If the CARMA process is minimum phase, this basically means that we can uniquely determine the values of the input white noise process from the output CARMA process. 

A stationary CARMA(p,q) process has the PSD
\begin{equation}
	P(f)  = \sigma^2 \frac{\left| \sum_{j=0}^q \beta_j (2\pi i f)^j \right|^2}{\left| \sum_{k=0}^p \alpha_k (2\pi i f)^k \right|^2}
	\label{eq-carma_psd}
\end{equation}
and autocovariance function at lag $\tau$
\begin{equation}
	R(\tau) = \sigma^2 \sum_{k=1}^p \frac{ \left[\sum_{l=0}^q \beta_l r_k^l \right] \left[\sum_{l=0}^q \beta_l (-r_k)^l \right] \exp(r_k \tau) }
		{ -2 \operatorname{Re}(r_k) \prod_{l=1, l \neq k}^p (r_l - r_k)(r^*_l + r_k) }.
	\label{eq-carma_autocovar}
\end{equation}
Here, $\operatorname{Re}(\cdot)$ denotes the real part and $z^*$ is the complex conjugate of $z$. In this work we only deal with the case when the roots are unique, as this is required in order to use the Kalman Filter to efficiently calculate the likelihood (see \S~\ref{s-likelihood}).

Most previous work on using continuous-time autoregressive processes for characterizing astronomical time series has focused on the case when $p = 1$ and $q = 0$, i.e., a CAR(1) model. The CAR(1) model is also called an Ornstein-Uhlenbeck process, and has often been referred to as a `damped random walk' in the astronomical literature. Using the notation above, the PSD and autocovariance for the more familiar case of $p=1,q=0$ are, respectively,
\begin{equation}
	P(f) = \sigma^2 \frac{1}{\alpha_0^2 + (2\pi f)^2}
	\label{eq-psd_car1}
\end{equation}
and
\begin{equation}
	R(\tau) = \frac{\sigma^2}{2\alpha_0} e^{-\alpha_0 \tau}.
	\label{eq-autocovar_car1}
\end{equation}
As can be seen, the PSD for the CAR(1) process is a Lorentzian function centered at zero with a break frequency at $\alpha_0 / (2\pi)$, while the autocorrelations decay exponentially with an $e$-folding time scale $1 / \alpha_0$. 

Inspection of the autocovariance function and PSD of a CARMA(p,q) process provides some guidance on how to interpret the CARMA(p,q) parameters. For the CARMA(p,q) process the autocorrelation function is a weighted sum of exponential functions, with the arguments of these exponential functions being the roots of the polynomial given by Equation (\ref{eq-characteristic}) and the weights being a function of the moving average coefficients. These roots may be complex-valued, although if $p$ is odd there is always at least one real root. As a result, the autocorrelation function for the CARMA(p,q) process is a sum of exponentially damped sinusoidal functions (corresponding to the complex roots) and exponential decays (corresponding to the real roots). The $e$-folding time scale of the decaying autocorrelations for each exponential function is $1.0 / |\operatorname{Re}(r_k)|$, while the frequencies of the oscillations in the autocorrelations are $|\operatorname{Im}(r_k)| / 2\pi$.

Because the PSD and the autocovariance function are a Fourier transform pair, we can also connect the roots of $A(\cdot)$ to the PSD. Because the Fourier transform of an exponentially damped sinusoidal function is a Lorentzian function, the PSD of a CARMA(p,q) process can be expressed as a weighted sum of Lorentzian functions. The widths of the Lorentzian functions are proportional to $|\operatorname{Re}(r_k)|$ while the centroids of the Lorentzian functions are given by $|\operatorname{Im}(r_k) / 2\pi|$. As with the autocovariance function, the moving average coefficients $\beta_1, \ldots, \beta_q$ help control the weights in the sum. Incidentally, a sum of Lorentzian functions is a common model used to characterize the X-ray PSDs of X-ray binaries \citep[e.g.,][]{Nowak2000a,Belloni2010a}. 

\subsection{State Space Representation}
\label{s-statespace}

The solution to Equation (\ref{eq-carma}) may be obtained by introducing a state space representation of a CARMA(p,q) process \citep[e.g.,][]{Brockwell2002a}. In addition, as discussed in Section \ref{s-likelihood}, representing a CARMA(p,q) process in a state space representation enables efficient calculation of the likelihood function for a measured time series. A state space representation models a stochastic process as arising from an observation equation and a state equation. The observation equation relates the observed time series to an unknown latent state variable, and the state equation describes the evolution of the state variable. Note that the state variable will in general not be scalar-valued. For the CARMA(p,q) model, the $p$-element state vector ${\bf x}(t)$ is a vector containing the value of a latent process $s(t)$ and its derivatives as a function of time $t$:
\begin{equation}
	{\bf x}(t) = \begin{pmatrix}
	s(t) \\
	s'(t) \\
	s''(t) \\
	\vdots \\
	s^{(p-1)} (t)
	\end{pmatrix}
	\label{eq-state_vector}.
\end{equation}
The state space representation of a CARMA(p,q) process $y(t)$ is
\begin{align}
	y(t) & = {\bf b} {\bf x}(t), \label{eq-obs_equation} \\
	d{\bf x}(t) & = A {\bf x}(t) dt + {\bf e} dW(t), \label{eq-state_equation}
\end{align}
where $W(t)$ is a Wiener process\footnote{The Wiener process is referred to as a standard Brownian motion.}, $dW(t)$ is a white noise process with mean zero and unit variance, ${\bf b} = [\beta_0, \beta_1, \ldots, \beta_{p-1}]$ is a $p$-element row vector, $\beta_j = 0$ for $j > q$, ${\bf e} = [0, 0, \ldots, 0, \sigma]^T$ is a $p$-element column vector, and $A$ is a $p \times p$ matrix with elements
\begin{equation}
	A = \begin{pmatrix}
		0 & 1 & 0 & \ldots & 0 \\
		0 & 0 & 1 & \ldots & 0 \\
		\vdots & \vdots & \vdots & \ddots & \vdots \\
		0 & 0 & 0 & \cdots & 1 \\
		-\alpha_0 & -\alpha_1 & -\alpha_2 & \cdots & \alpha_{p-1}
		\end{pmatrix}.
		\label{eq-A_matrix}
\end{equation}

The solution to Equation (\ref{eq-state_equation}) with random initial condition ${\bf x}(t=0) = {\bf x}_0$ is \citep[e.g.,][]{Brockwell2002a}
\begin{equation}
	{\bf x}(t) = e^{At} {\bf x}_0 + \int_{0}^{t} e^{A(t-u)} {\bf e} \ dW(u).
	\label{eq-carma_solution}
\end{equation}
The first term on the right hand side represents the deterministic contribution to the evolution of the state vector, given the initial condition, while the second term on the right hand side is a random variable representing the stochastic contribution to this evolution. Note that because $dW(u)$ has an expectation value of zero, the stochastic integral in Equation (\ref{eq-carma_solution}) also has an expectation value given by the zero vector.
 
The process given by the solution to Equation (\ref{eq-carma_solution}) is stationary if and only if ${\bf x}(0)$ has an expectation value given by the zero vector and $p \times p$ covariance matrix with elements \citep{Jones1990a}
\begin{equation}
	V_{ij} = -\sigma^2 \sum_{k=1}^p \frac{r_k^i (-r_k)^j}{2 \operatorname{Re}(r_k) \prod_{l=1, l \neq k}^p (r_l - r_k) (r_l^* + r_k)}.
	\label{eq-state_covar}
\end{equation}
In this case, the stationary mean of ${\bf x}(t)$ is also the zero vector and the stationary covariance is also given by Equation (\ref{eq-state_covar}). Because ${\bf x}(t)$ is stationary, the CARMA process defined by Equation (\ref{eq-obs_equation}) is also stationary with mean zero and variance ${\bf b} V {\bf b}^T$; note that this is the same as that given by Equation (\ref{eq-carma_autocovar}) using a time lag of $\tau = 0$.

\section{Statistical Inference for CARMA Models}
\label{s-inference}

If the white noise term in Equation (\ref{eq-carma}) is Gaussian, the resulting CARMA(p,q) process is also Gaussian. In this case, the likelihood function for a CARMA(p,q) model may be derived for a measured time series ${\bf y} = [y_1,\ldots,y_n]^T$ sampled at times $t_1,\ldots,t_n$  as
\begin{align}
	p({\bf y}|\mu,\sigma,\alpha,\beta)& \propto \frac{1}{|\Sigma|} \exp \left \{ -\frac{1}{2} ({\bf y} - \mu)^T \Sigma^{-1} ({\bf y} - \mu) \right \}, \label{eq-likhood_slow} \\
	\Sigma_{ij}& = R(|t_i - t_j|) + \delta_{ij} \sigma^2_{i}, \label{eq-likhood_covar}
\end{align}
where $\mu$ is the mean of the time series, $\alpha = (\alpha_0,\ldots, \alpha_{p-1}), \beta = (\beta_1, \ldots, \beta_q)$, $\delta_{ij}$ is the Kronecker delta function, $\sigma^2_i$ is the variance in the measurement error for $y_i$, and $R(\cdot)$ is the autocovariance function of a CARMA(p,q) process, given by Equation (\ref{eq-carma_autocovar}). Note that here and throughout this work we assume that the measurement errors on the time series are uncorrelated. Maximum-likelihood estimates of the parameters $\mu, \sigma, \alpha,$ and $\beta$ may be obtained by maximizing Equation (\ref{eq-likhood_slow}), and Bayesian inference may be performed by combining Equation (\ref{eq-likhood_covar}) with a suitably chosen prior.

Calculating Equation (\ref{eq-likhood_slow}) directly requires inverting the $n \times n$ covariance matrix $\Sigma$, the computational complexity of which scales as $O(n^3)$. This may represent a considerable bottleneck for large time series, especially when performing statistical inference for a large sample of objects. Fortunately, the state space representation of a CARMA(p,q) process enables application of the Kalman Filter, which speeds up calculation of the likelihood function to $O(n)$ operations.

\subsection{Likelihood Calculation via the Kalman Recursions}
\label{s-likelihood}

For state space models of Gaussian processes, such as that described by Equations (\ref{eq-obs_equation})--(\ref{eq-state_equation}), the Kalman filter algorithm may be used to sequentially and efficiently calculate the mean and covariance of the next state and observation of a sampled process given the previous values and the model parameters \citep[e.g.,][]{Brockwell2002a}. Because a normal distribution is completely characterized by its mean and covariance, this is all that we need to calculate the likelihood function for a measured time series. 

We factor the likelihood as
\begin{equation}
	p({\bf y}|\mu,\sigma,\alpha,\beta) = p(y_1|\mu,\sigma,\alpha,\beta) \prod_{i=2}^n p(y_i|y_1,\ldots,y_{i-1},\mu,\sigma,\alpha,\beta).
	\label{eq-factored_likhood}
\end{equation}
For a Gaussian CARMA(p,q) model each of the terms on the right hand side of Equation (\ref{eq-factored_likhood}) are normal distributions:
\begin{align}
	p({\bf y} | \mu, \sigma, \alpha, \beta)& \propto \prod_{i=1}^n \frac{1}{Var(y_i|{\bf y}_{<i}, \sigma, \alpha, \beta)} \nonumber \\
	& \times \exp \left \{ -\frac{1}{2} \frac{(y_i - E(y_i | {\bf y}_{<i}, \mu, \sigma, \alpha, \beta))^2}{Var(y_i|{\bf y}_{<i}, \sigma, \alpha, \beta)} \right \}.
	\label{eq-likhood}
\end{align}
Here we have used the notation ${\bf y}_{<i} = [y_1, \ldots, y_{i-1}]$. The Kalman Filter is used to calculate the means and variances for each of these normal distributions, efficiently calculating the likelihood in $O(n)$ operations. 

Because astronomical time series are measured with error, we introduce a measurement error term to the observation equation of the state space representation, modifying Equation (\ref{eq-obs_equation}) to become
\begin{equation}
	{\bf y}(t) = {\bf b} {\bf x}(t) + \delta(t).
	\label{eq-obs_equation2}
\end{equation}
The measurement error for $y_i$, $\delta(t_i)$, is assumed to be normally distributed with mean zero and variance $\sigma_{i}^2$. Equations (\ref{eq-obs_equation2}) and (\ref{eq-state_equation}) are a state space representation for a time series measured with error assuming a CARMA(p,q) model.

In order to use the Kalman Filter for a sampled continuous time process, it is necessary to convert the continuous time state space representation to that of a discrete time process evaluated at the sampled time values. This requires integrating the state equation (Eq.[\ref{eq-state_equation}]) over the time intervals between subsequent observations. The discrete state space representation of a CARMA process sampled at times $t_1,\ldots,t_i,\ldots,t_n$ is derived using Equation (\ref{eq-carma_solution}) \citep[e.g.,][]{Jones1990a} to be
\begin{align}
	y_i& = {\bf b x}_i + \delta_i \label{eq-discrete_obs_equation} \\
	{\bf x}_i& = e^{A(t_i - t_{i-1})} x_{i-1} + \eta_i \label{eq-discrete_state_equation}
\end{align}
where $\delta_i$ denotes the normally distributed measurement error at $t_i$ with mean zero and variance $\sigma_i^2$ and
\begin{equation}
	\eta_i = \int^{t_i}_{t_{i-1}} e^{A(t_i - t)} {\bf e}\ dW(t)
	\label{eq-discrete_error}
\end{equation}
is a random vector drawn from a multivariate normal distribution with mean given by the zero vector and covariance matrix \citep[e.g.,][]{Gardiner2004a}
\begin{equation}
	Cov(\eta_i) = \int_{0}^{t_i - t_{i-1}} e^{At} {\bf e} {\bf e}^T e^{A^Tt}\ dt.
	\label{eq-discrete_error_covar}
\end{equation}
Equations (\ref{eq-discrete_obs_equation})--(\ref{eq-discrete_error_covar}) provide everything we need to compute the Kalman Filter and the likelihood function for a CARMA model.

Calculation of the matrix exponentials needed in Equations (\ref{eq-discrete_state_equation}) and (\ref{eq-discrete_error_covar}) is computationally expensive. However, for diagonal matrices the matrix exponential is trivial to calculate. In order to improve the efficiency of the Kalman Filter we use the diagonal form $A = U D U^{-1}$ \citep{Jones1990a}, where the columns of $U$ contain the right eigenvectors of $A$ and $D$ is a diagonal matrix:
\begin{align}
	U_{lk}& = r^{l-1}_k \label{eq-eigenmatrix} \\
	D_{lk}& = \begin{cases}
		r_k & l = k \\
		0 & l \neq k
		\end{cases}.
		\label{eq-transition_matrix}
\end{align}
Recall that $r_k, k=1, \ldots, p$ are the roots of the autoregressive polynomial. We then transform Equations (\ref{eq-discrete_obs_equation}) and (\ref{eq-discrete_state_equation}) to be in terms of the rotated state vectors $\tilde{\bf x}_i = U^{-1} {\bf x}_i$:
\begin{align}
	y_i& = \tilde{\bf b} \tilde{\bf x}_i + \delta_i \label{eq-rot_obs_equation} \\
	\tilde{\bf x}_i& = \Lambda_i \tilde{\bf x}_{i-1} + \tilde{\eta}_i. \label{eq-rot_state_equation}
\end{align}
Here $\Lambda_i$ is a diagonal matrix with diagonal elements $\Lambda_{i,kk} = e^{r_k(t_i - t_{i-1})}$, and $\tilde{\bf b} = {\bf b} U$. The stochastic term in the rotated state space formulation, $\tilde{\eta}_i$, is a complex valued normally distributed random variable with mean given by the zero vector and Hermitian covariance matrix. The elements of the covariance matrix of $\tilde{\eta}_i$ are not actually needed for the Kalman Filter \citep[e.g.,][]{wang2013cts} in our implementation, but they are given in \citet{Jones1990a}. We note that the variables in the rotated state space notation are complex-valued, although they are all real-valued in the original representation. 

\citet{Jones1981a} derived the Kalman Filter for a CAR(p) model under the rotated state representation, and \citet{Jones1990a} extended these results to a CARMA(p,q) model. In \S~\ref{s-kfilter} of the Appendix we provide the Kalman filter algorithm for a CARMA($p,q$) model, and refer the reader to \citet{Jones1981a} and \citet{Jones1990a} for further details. After running the Kalman filter, we will have the values of $E(y_i|{\bf y}_{<i},\theta)$ and $Var(y_i|{\bf y}_{<i},\theta)$ needed for computing the likelihood function via Equation (\ref{eq-likhood}).

\subsection{Interpolation and Extrapolation from a Measured Time Series}
\label{s-interpolation}

In certain applications one may need to simulate an interpolated or extrapolated time series conditional on a measured time series. Example of this include reverberation mapping of AGN \citep{Horne1991a,Zu2011a,Brewer2011a,Pancoast2011a,Pancoast2012a} or studies of the time delay between images of gravitationally lensed quasars \citep[e.g.,][]{Press1992a,Kochanek2004a,Morgan2012a}. In addition, forecasting may also be useful for generating alerts. The probability distribution of a Gaussian CARMA process at time $t_0$ given the measured time series is a normal distribution, and one can use the usual Gaussian Process machinery to calculate the conditional mean and variance of future data. However, as with the likelihood calculation this is expensive, also requiring an expensive matrix inversion. Instead, for a CARMA process we can use the Kalman Filter to efficiently calculate the mean and variance of this normal distribution, and in this section we derive these quantities.

Denote the value of a CARMA process at time $t_0$ as $y_0$, and define $j(t_0) = \min \{i;  t_0 < t_i \}$, i.e., $t_{j(t_0)-1} < t_0 < t_{j(t_0)}$. Assuming a Gaussian CARMA process, we can write the probability distribution of $y_0$ given a CARMA model and a measured time series ${\bf y}$ as
\begin{multline}
	p(y_0|{\bf y}, \theta) \propto \\ 
	\quad \frac{1}{Var(y_0|{\bf y}_{<j(t_0)},\theta)} \exp \left \{-\frac{1}{2}\frac{(y_0 - E(y_0|{\bf y}_{<j(t_0)}, \theta))^2}{Var(y_0|{\bf y}_{<j(t_0)}, \theta)}\right \} \\
	\quad \times \prod_{i=j(t_0)}^n \frac{1}{Var(y_i|y_0,{\bf y}_{<i}, \theta)} \exp \left \{ -\frac{1}{2} \frac{(\tilde{y}_i - \tilde{c}_i - \tilde{d}_i \tilde{y}_0)^2}{Var(y_j|y_0,{\bf y}_{<i}, \theta)} \right \}.
\label{eq-linear_filter}
\end{multline}
As before, $\tilde{y}_i = y_i - \mu$. The values of $Var(y_0|{\bf y}_{<j(t_0)},\theta), Var(y_i|y_0,{\bf y}_{<i},\theta),$ and $E(y_0|{\bf y}_{<j(t_0)},\theta)$ can be obtained by running the Kalman filter on the time series generated by inserting $t_0$ into the set of observation times $t_i$. The sets of coefficients $\tilde{c}_i$ and $\tilde{d}_i$ can be obtained recursively through an algorithm similar to the Kalman filter, which we describe in \S~\ref{s-coefs} of the Appendix. 

From Equation (\ref{eq-linear_filter}) we can derive the mean and variance of $y_0$ as
\begin{align}
	E(y_0|{\bf y},\theta) & = Var(y_0|{\bf y},\theta) \nonumber \\
	& \times \left [ \frac{E(y_0|{\bf y}_{<j(t_0)},\theta)}{Var(y_0|{\bf y}_{<j(t_0)},\theta)} \right. \nonumber \\
	& + \left. \sum_{i=j(t_0)}^n \frac{\tilde{d}_i (\tilde{y}_i - \tilde{c}_i)}{Var(y_i|y_0,{\bf y}_{<i}, \theta)} \right] \label{eq-predicted_cmean} \\
	Var(y_0|{\bf y},\theta) & = \left [ \frac{1}{Var(y_0|{\bf y}_{<j(t_0)},\theta)} \right. \nonumber \\
	& \left. + \sum_{i=j(t_0)}^n \frac{\tilde{d}^2_i}{Var(y_i|y_0,{\bf y}_{<i}, \theta)} \right]^{-1} \label{eq-predicted_cvar}.
\end{align}
Equations (\ref{eq-predicted_cmean}) and (\ref{eq-predicted_cvar}) provide the interpolated or extrapolated value and its uncertainty, assuming the CARMA model. Note that for forecasting and backcasting, the first and second terms appearing in the right hand sides of Equations (\ref{eq-predicted_cmean}) and (\ref{eq-predicted_cvar}) are ignored, respectively.

An interpolated or extrapolated time series may be simulated sequentially at time values $\hat{t}_1, \ldots, \hat{t}_m$ by first using Equations (\ref{eq-predicted_cmean}) and (\ref{eq-predicted_cvar}) to generate a value of $y(\hat{t}_1)$ from a normal distribution, inserting this value of $y(\hat{t}_1)$ into the measured time series array, and repeating for the remaining $\hat{t}_j$.

\subsection{Bayesian Inference}
\label{s-bayes}

In this work we focus on Bayesian inference of the CARMA process model. This is primarily because in Bayesian inference one derives the probability distribution of the CARMA process given the measured time series (i.e., the `posterior' distribution), providing a rigorous assessment of the uncertainties in the CARMA model, and consequently in the inferred power spectrum. The likelihood function of a CARMA model can exhibit multiple maxima for $p > 1$ \citep[e.g.][]{Broersen2006a}, and therefore traditional techniques based on the Fisher information matrix and the asymptotic normality of the maximum likelihood estimate do not apply in general for CARMA models.

Bayesian inference is based on the posterior distribution, which is related to the likelihood function by the equation
\begin{equation}
	p(\theta|{\rm y}) \propto p(\theta) p({\bf y}|\theta) \label{eq-posterior}.
\end{equation}
Here, $p(\theta)$ is the prior distribution on the model parameters. Because we have already derived the likelihood function, the only additional thing we need to do to perform Bayesian inference is to specify the prior distribution. In our model we assume a uniform prior on the standard deviation of the lightcurve subject to $[R(0)]^{1/2} < R_0$ for some input value of $R_0$ and a uniform prior on $\mu$. In this work we set $R_0$ to be ten times the standard deviation of the measured time series.

Following \citet{Jones1981a}, we use the following parameterization for $\alpha$:
\begin{multline}
A(z) = (a_1 + a_2 z + z^2) (a_3 + a_4 z + z^2) \cdots \\
	\times \begin{cases}
	(a_{p-1} + a_p z + z^2) & \text{if $p$ is even}, \\
	(a_p + z) & \text{if $p$ is odd}.
	\end{cases}
	\label{eq-alpha_parameterization}
\end{multline}
The roots of the autoregressive polynomial will have negative real parts (and thus produce a stationary CARMA process) if and only if $a_1, \ldots, a_p$ are positive. In order to enforce this, we sample the values of $\log a_k$ in our MCMC sampler, and place a uniform prior on their values. Using the parameterization of Equation (\ref{eq-alpha_parameterization}) also has the computational convenience that the roots may be analytically computed from the quadratic and linear terms. In addition, we use a similar parameterization and prior for the moving average coefficients, $\beta$. This parameterization of the moving average coefficients enforces the system to be minimum phase by keeping the roots of the moving average polynomial positive. 

Because the likelihood function is invariant to permutations of the indices of $r_k$, we place the following ordering constraint on the indices in order to make the model identifiable:
\begin{equation}
	\operatorname{Im}(|r_1|) > \operatorname{Im}(|r_3|) > \ldots > \operatorname{Im}(|r_p|) \label{eq-roots_prior}
\end{equation}
The constraints are only with respect to the pairs $(r_k, r_{k+1})$ for $k$ odd because the roots come in complex conjugate pairs. Finally, we also constrain the Lorentzian centroid values to be less than the inverse of the minimum time between measurements, and the Lorentzian widths to be between the minimum and maximum frequencies probed by the light curve time sampling.

In practice, we include an additional scaling parameter on the measurement errors, $\nu$, such that the true measurement error variances are assumed to be $\varsigma_i^2 = \nu \sigma^2_i$. Because the derived PSD depends on the amplitude of the measurement errors, especially at the high-frequency end, we include this additional parameter to incorporate uncertainty on the quoted measurement error variances $\sigma^2_i$. Our prior on the parameter scaling parameter $\nu$ is a scaled inverse $\chi^2$ distribution with 50 degrees of freedom and scale parameter of unity. In addition, we bound the value to be $1/2 < \nu < 2$. This prior reflects an assumption that the relative amplitude of the measurement errors are correct, but that the overall normalization of the measurement error standard deviations $\sigma_i$ has an uncertainty of $10\%$. This is application dependent, and researchers analyzing time series for which they have greater confidence in the quoted measurement error amplitudes may wish to use a much larger value than 50 degrees of freedom, or narrower bounds on $\nu$.

\subsection{Assessing the Fit}
\label{s-assess_fit}

The quality of the fit, and the appropriateness of the Gaussian CARMA model, can be assessed by noting that the standardized residuals, $\chi_i$, from the Kalman filter should have the form of Gaussian white noise. The standardized residuals are calculated as
\begin{equation}
	\chi_i = \frac{y_i - E(y_i|{\bf y}_{<i}, \hat{\theta})}{[Var(y_i|{\bf y}_{<i}, \hat{\theta})]^{1/2}},
	\label{eq-standard_resids}
\end{equation}
where $\hat{\theta}$ is a point estimate of $\theta$. When inspecting the residuals we have found that it is best to use for $\hat{\theta}$ the value of $\theta$ obtained from maximizing the posterior or likelihood. As discussed in \S~\ref{s-computation}, the posterior for the CARMA model parameters can be multi-modal, especially for high values of $p$ or $q$, and the CARMA process implied by the posterior median or mean may not be representative if these quantities fall between the modes.

If the Gaussian CARMA model is correct, then the residuals should have a normal distribution with mean zero and standard deviation of unity. Deviations from the expected normal distribution can be used to assess the assumption of a Gaussian process. Similarly, the sequence $\chi_1, \ldots, \chi_n$ should form a Gaussian white noise sequence, the accuracy of which can be assessed through the autocorrelation function of the standardized residuals. Under the null hypothesis that the sequence $\chi_1,\ldots,\chi_n$ are a white noise sequence, their sample autocorrelations are approximately independently and normally distributed with mean zero and variance $1 / n$. If a large number of the sample autocorrelations are outside of the, say, $2\sigma$ interval then there is residual correlation structure in the time series that the CARMA models is not picking up. Finally, the sequence of squared residuals $\chi^2_1,\ldots,\chi^2_n$ should also form a white noise sequence under the assumption of a Gaussian CARMA model. Therefore, deviations of the sample autocorrelations of the sequence $\chi^2_1,\ldots,\chi^2_n$ from a zero mean normal distribution with variance $1/n$ signal non-linear behavior, as they indicate that the variance in the time series is changing with time. Note that in both these cases it is the ACF of the sequence of residuals that is calculated, and not the time series of residuals. In other words, we calculate the ACF of the residuals treating their time values as being on a regular grid: $t = 1, 2, ..., n-1, n$.

\subsection{Choosing the Order of the CARMA Model}
\label{s-order}

There are multiple approaches to choosing the order of the CARMA model. First, it is often the case that traditional methods based on statistical hypothesis testing are not optimal for flexible models such as the CARMA model. For one, CARMA models are not nested because a CAR($p-1$) model cannot be obtained by setting $\alpha_p$ in a CAR($p$) model to some finite value.  Therefore, the usual asymptotic assumptions underlying the likelihood ratio test do not apply. An exception is the transformed CAR model presented by \citet{Belcher1994a}, which does provide a sequence of nested models. Second, choosing the order of the CARMA model is more of a model \emph{selection} issue, and should not be framed within the context of ruling out a null hypothesis. For many applications the CARMA parameters will not have any specific physical meaning, so there will not be any physically meaningful null hypothesis. Instead, in time series analysis it is common to choose the order of the model to be that which best predicts additional data (i.e., minimizes the test error), or otherwise is `close' to the process that generated the data. Because we are interested in using the CARMA model to adequately and flexibly constrain the PSD and correlation structure in the time series, as well as to provide automatic variability features that may be used, e.g., for classification, our approach is to choose the $p$ and $q$ that minimize an estimate of how close the CARMA($p,q$) model is to the data generating process.

Information criteria are a common mechanism for ranking a set of models. Such criteria are often used as approximations to the prediction error of future data, and are usually inexpensive to calculate. In time series analysis it is common to use the Akaike Information Criterion \citep[AIC,][]{Akaike1973a}, which is based on the maximum-likelihood estimate of $\theta$. The AIC provides an estimate of the relative information lost in using a model to represent the underlying process that generated the data, as measured through the Kullback-Leibler divergence. The AIC in its original form is strictly only valid asymptotically, but \citet{Hurvich1989a} provide a correction to AIC for finite sample sizes (denoted as AICc). An alternative Bayesian criteria is the Deviance Information Criterion \citep[DIC,][]{spiegelhalter2002bayesian}, which may be calculated from the samples returned by a MCMC sampler. Cross-validation is another approach for choosing $p$ and $q$, as it provides an estimate of the test error. Cross-validation works by dividing the light curve into $K$ contiguous subsamples, fitting the model to the data after removing one of the subsamples, evaluating the model performance on the withheld subsample, and repeating the for the remaining $K-1$ subsamples. The errors on the withheld subsamples are averaged, and $(p,q)$ can be chosen to minimize this error. 

Because it is more expensive to run independent MCMC samplers than to obtain a maximum-likelihood estimate for each set of candidate $(p,q)$ values, we use the AICc instead of DIC to choose the order of the CARMA model. Similarly, because it is expensive to perform a fit to each of the $K$ subsamples generated by cross-validation we use AICc in this work.  The AIC is defined as
\begin{equation}
AIC(p,q) = 2 k - 2 \log p({\bf y}|\hat{\theta}_{\rm mle}, p, q),
\label{eq-aic}
\end{equation}
where $k$ is the number of parameters in the CARMA($p,q$) model, and $\hat{\theta}_{\rm mle}$ is the maximum-likelihood estimate of the CARMA process parameters, $\theta$. The best model is the one that minimizes the AIC. The AIC penalizes against overfitting through the $2k$ term: once the improvement to the log-likelihood function that results from using a more complex model does not increase faster than the number of parameters, the AIC will begin to worsen. The AICc is
\begin{equation}
AICc(p,q) = AIC(p,q) + \frac{2k(k+1)}{n - k - 1},
\label{eq-aicc}
\end{equation}
where $n$ is the number of data points in the lightcurve. The AICc places a stronger penalty for model complexity due to the finite sample correction.

Finally, we note that in certain cases it is scientifically meaningful to assess the significance of a specific feature in the PSD. For example, CARMA models selected based on AICc to have $p > 1$ may show evidence for quasi-periodic oscillation (QPO) features in the PSD. These features in, for example, light curves from black hole systems are thought to be driven by astrophysical phenomenon in accretion flows \citep[e.g.,][]{Done2007a}, and therefore in this case there is a scientifically meaningful null hypothesis. For such situations one needs to assess the statistical significance of these features, even if their existence provides a better AICc. This can be done by inspecting the posterior distribution of the feature of interest for the chosen model (e.g., the one with the minimum AICc). To use the QPO example, the coherence of a QPO is often quantified via a `quality factor', which is the ratio of peak frequency of the QPO to its width. The `statistical significance', or, more importantly, the scientific significance of a possible QPO feature could be assessed by inspecting the posterior distribution of the QPO quality factor. If most of the posterior probability is at high quality factors, then one may be confident in the existence of this QPO (assuming the CARMA model has been shown to be accurate, see \S~\ref{s-assess_fit}). Otherwise, if most of the probability in the quality factor is at low values, then this `QPO' feature may not be scientifically meaningful even if its inclusion still provides a more accurate model.

\subsection{Computational Considerations and Software}
\label{s-computation}

The computational complexity of evaluating the likelihood function using the Kalman filter scales as $O(n)$ for a time series with $n$ data points. Unfortunately, because the Kalman filter is a serial calculation it cannot be parallelized. 

For our MCMC sampler we use the robust adaptive Metropolis algorithm of \citet{Vihola2012a} with a Student's $t$-distribution with eight degrees of freedom as the proposal distribution. The algorithm of \citet{Vihola2012a} improves upon the standard Metropolis algorithm by adaptively tuning the covariance matrix of the proposals to achieve a desired acceptance rate; in this work we use an acceptance fraction of 25\%. We only adapt the proposals during the burn-in phase\footnote{MCMC samplers are first run for a `burn-in' phase in order to forget about how the MCMC sampler was started. The sampled parameter values are not saved until the burn-in phase has finished, at which point the MCMC should have converged to the posterior distribution.}.

For $p > 1$ the likelihood function often contains multiple modes, especially for higher orders of $p$ and $q$. This presents a difficulty for many optimizers and MCMC samplers. When computing the maximum-likelihood estimates we run a local optimization algorithm using 100 random starting values of $\theta$, and choose the best $\hat{\theta}_{\rm mle}$ among the outputs. While there is no guarantee that this approach will find the global optimum, we have found it to be sufficient for the purposes of choosing the values of $p$ and $q$ via the AICc. Further improvement may be obtained through the use of, for example, genetic algorithms. 

In order to effectively sample the posterior for $p > 1$, we also employ parallel tempering in our MCMC sampler \citep[e.g.,][]{Liu2004a}. In our parallel temperating implementation $K$ parallel chains are run using their own robust adaptive Metropolis algorithm, where the $k^{\rm th}$ chain samples from the distribution $p(\theta|{\rm y})^{1/T_k}$, and the sequence $T_1 < \ldots < T_K$ forms what is known as a `temperature ladder'; note that $T_1 = 1$. Denote the value of the CARMA model parameters for the $k^{\rm th}$ chain as $\theta_k$. After each chain updates their parameters via the \citet{Vihola2012a} Metropolis step, we then propose to swap values the values of $\theta_k$ and $\theta_{k-1}$ for $k = K, \ldots, 2$. The purpose of the temperature ladder is to flatten the posterior distribution for larger values of $T_k$, enabling the chains to move between modes in the `hot' chains. The swapping step then allows the coolest chain, which is the one we actually care about, to jump between modes when the hotter chains find these modes. We use a temperature ladder that forms a regular grid in $\log T_k$, with $T_K = 100$. In general we have found that values of $K = 10$ are sufficient. Although we do not do so in our code, the parallel tempering algorithm is parallelizable. Further details on parallel tempering, and MCMC methods in general can be found in \citet{Liu2004a}.

We have developed software to run our MCMC algorithm on a time series, assuming the CARMA model. The software is written in a combination of C++ and Python, with the C++ code forming a Python extension. The MCMC sampler is written in C++ for speed, but may be called from within Python, enabling one to analyze the results within Python. The Python component of our software also contains a routine for computing the maximum-likelihood estimate of $\theta$ (although the likelihood calculations are done in C++), choosing the order of the CARMA model via AICc, and routines for analyzing the MCMC output. Our software is available at \url{https://github.com/bckelly80/carma\_pack}.

\section{Example Applications to Simulated Time Series}
\label{s-simulations}

In order to illustrate the use of CARMA models for analysis of astronomical time series, we generated a mock light curve from a CARMA(5,3) process under both regular and irregular sampling, and a non-stationary irregularly sampled light curve that switches from one CARMA(5,3) process to another. 

\subsection{Stationary Process Under Regular Sampling}
\label{s-stationary_regular}

\begin{figure}
	\includegraphics[scale=0.45,angle=00]{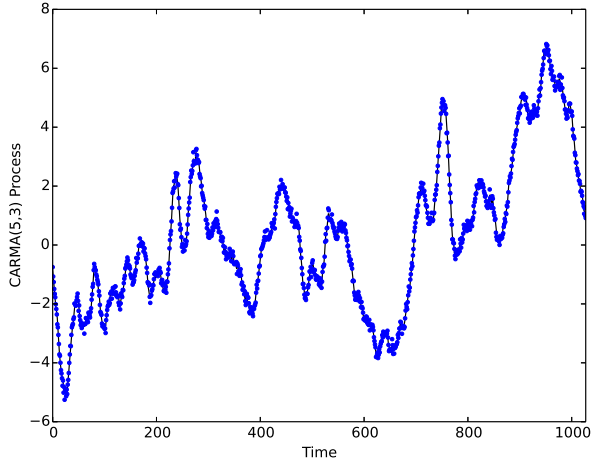}
	\caption{Simulated light curve from a CARMA(5,3) process on a regular grid. The black line denotes the true values, and the blue dots denote the measured values. Because the measurement errors are very small for this light curve, the measured values track the true values very closely. \label{f-mock1_lcurve}}
\end{figure}

\begin{figure}
	\includegraphics[scale=0.45,angle=00]{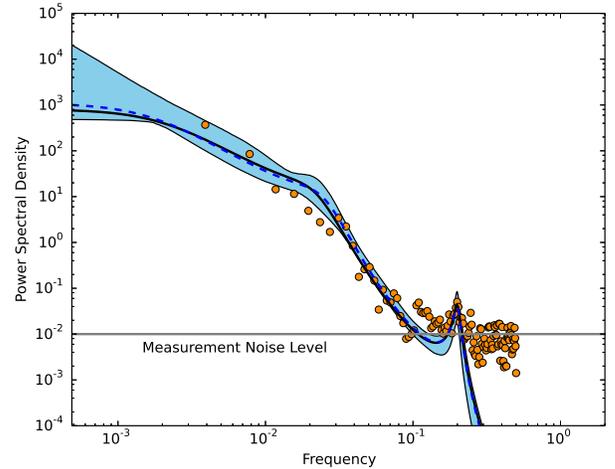}
	\caption{Power spectral density for the light curve shown in Figure \ref{f-mock1_lcurve}. The true PSD is given by the solid black line, the periodogram by the orange circles, the PSD from the maximum-likelihood estimate assuming a CARMA(5,1) model (chosen to minimize AICc) by the blue dashed line, and the blue region contains $95\%$ of the probability on the PSD assuming a CARMA(5,1) model. There is a weak oscillatory feature centered at a frequency of $1 / 5\ {\rm day}^{-1}$ which is at the measurement noise level. This feature is not obvious above the measurement error component for the periodogram, but the CARMA model is able to recover it, along with the rest of the PSD. We note that the tight errors on the PSD below the measurement noise level are due to extrapolation assuming the parametric form of CARMA(5,1) model, and using a higher order model would enable more flexibility, and consequently broader errors below the measurement noise level.
	 \label{f-mock1_psd}}
\end{figure}

For the first light curve we simulated a CARMA(5,3) process sampled on a regular grid $t_1 = 1, \ldots, t_n = 1028$ days. The CARMA(5,3) parameters were chosen to ensure that the mock light curve was dominated by broad-band noise, as with, for example AGN. The PSD for this light curve is flat on time scales $\gtrsim 500$ days, falls off as $\sim 1 / f^2$ for frequencies corresponding to time scales $500 \lesssim \Delta t \lesssim 50$ days, and steepens to $\sim 1 / f^4$ for frequencies corresponding to time scales $\lesssim 50$ days. In addition, there is a weak QPO centered at a frequency of $\sim 1 / 5\ {\rm day^{-1}}$. The measurement noise level for this source was chosen to be just below the magnitude of the QPO, in order to test if we can recover an oscillatory feature at the limit of the measurement noise. The mock light curve is shown in Figure \ref{f-mock1_lcurve}, and the PSD for this light curve is shown in Figure \ref{f-mock1_psd}. Note that the measurement error in this case is only $\approx 1\%$ of the standard deviation in the light curve. In addition, because the light curve is regularly sampled we also show its periodogram in Figure \ref{f-mock1_psd}. The QPO feature appears but is not obvious above the measurement noise component of the periodogram, although one may suspect its existence through visual inspection.

\begin{figure}
	\includegraphics[scale=0.45, angle=00]{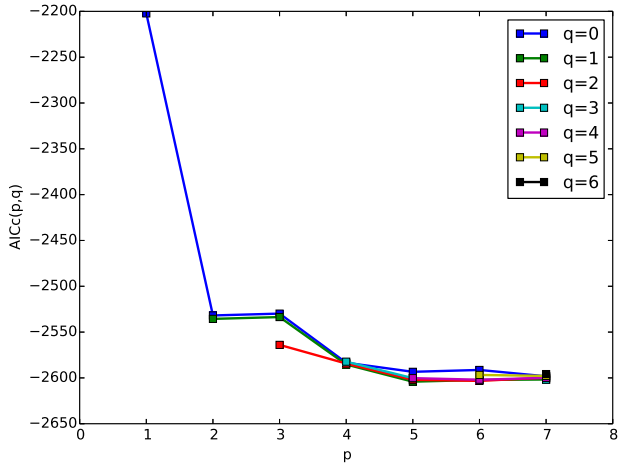}
	\caption{The AICc values computed from the simulated light curve shown in Figure \ref{f-mock1_lcurve} for CARMA($p,q$) models of order $p \leq 7, q < p$. The minimum AICc is achieved for the values $p=5,q=1$ although there is little change in the AICc for models of order $p \geq 5$.
	\label{f-mock1_aicc}}
\end{figure}

We searched for models of the form CARMA($p,q$) for $p = 1, \ldots, 7, q = 0, \ldots, p-1$. For each value of $(p,q)$ we computed a maximum-likelihood estimate, which we then used to compute the AICc. The values of AICc as a function of $p$ and $q$ are shown in Figure \ref{f-mock1_aicc}. The AICc drops off rapidly down to $p = 4$, after which it levels off. The minimum AICc is obtained for $p=5$ and $q=1$, although models with $p \geq 5$ provide comparable quality. As discussed in \S~\ref{s-order}, the fact that the AICc chose a model with $p=5, q=1$ even though the true model is $p=5, q=3$ implies that the difference in log-likelihood between the two models was not sufficiently large to warrant the inclusion of the additional parameters required under the CARMA(5,3) model. However, because the AICc depends on the data, it can change for different realizations from the same stochastic process. Because of this, different simulations of the CARMA(5,3) light curve may result in different values of $p,q$ for which the AICc is minimized.

Using the values of $p = 5, q = 1$, we ran our MCMC sampler using 10 parallel chains for $7.5 \times 10^4$ iterations, discarding the first $2.5 \times 10^4$ as burn-in. The residuals calculated using the best-fit CARMA(5,1) model were consistent with a unit variance Gaussian white noise sequence, suggesting that the CARMA(5,1) model provides an adequate fit. 

In Figure \ref{f-mock1_psd} we show the maximum-likelihood estimate of the model PSD and the region containing $95\%$ of the probability on the PSD. The chosen CARMA(5,1) model recovers the PSD, including the QPO feature, the centroid of which corresponds to an estimated time scale of $5.04 \pm 0.08$ days. Note that the tight constraints on the PSD below the noise level are caused by extrapolation of the CARMA(5,1) model form, and are not reflective of the actual uncertainty on the PSD in this regime when one does not know the order of the CARMA process. Because the PSD is largely unconstrained below the measurement noise level, the uncertainties would be larger in this regime if we had used a larger value of $p$. 

\subsection{Stationary Process Under Irregular Sampling}
\label{s-stationary_irregular}

\begin{figure}
	\includegraphics[scale=0.45,angle=00]{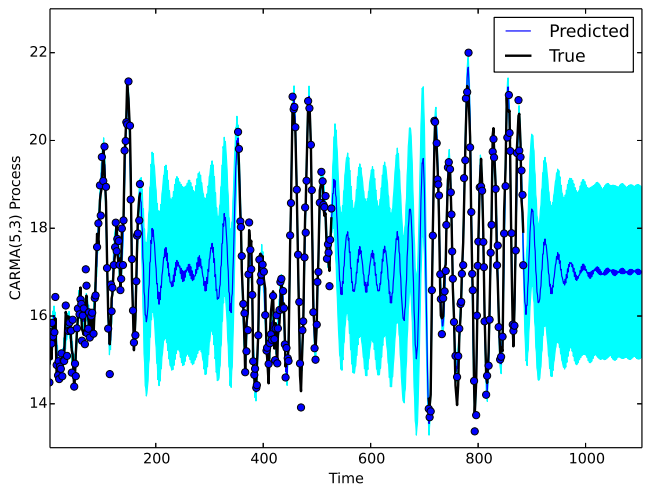}
	\caption{Simulated light curve from a CARMA(5,3) process irregularly sampled over three `observing seasons'. The black line denotes the true values, and the blue dots denote the measured values. Also shown are interpolated and forecasted values, based on the best-fitting CARMA(5,1) process; a CARMA(5,1) model had the minimum AICc value. The solid blue line and cyan region denotes the expected value and $1\sigma$ error bands of the interpolated and extrapolated light curve, given the measured light curve. \label{f-mock2_lcurve}}
\end{figure}

\begin{figure}
	\includegraphics[scale=0.45,angle=00]{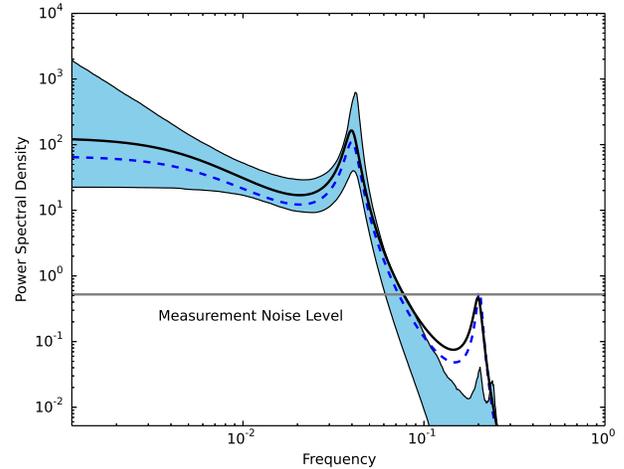}
	\caption{Power spectral density for the light curve shown in Figure \ref{f-mock2_lcurve}, with symbols the same as for Figure \ref{f-mock1_psd}. The constraints on the PSD are given by a $p = 5, q = 1$ model. A $p = 5, q = 1$ model was found to have the minimum AICc and is sufficient to capture the variability characteristics above the measurement noise.
	 \label{f-mock2_psd}}
\end{figure}

For our second simulated light curve, we used a CARMA(5,3) process but with different parameters, as well as a sampling pattern and measurement errors that are more realistic of an actual optical light curve. We simulated three observing seasons of 90 epochs separated by 180 days with time spacing drawn from a uniform probability distribution over 1 to 3 days. The measurement error standard deviations were set to $20\%$ of the standard deviation in the light curve. In this case we used a PSD that has a strong oscillator mode centered at a frequency of $1 / 25\ {\rm day}^{-1}$; this type of PSD is more representative of certain types of variable stars. As with the first simulated light curve, there is a weak oscillatory feature at $1 / 5\ {\rm day}^{-1}$, but in this case the feature falls primarily below the measurement noise level. The simulated light curve is shown in Figure \ref{f-mock2_lcurve}, and its PSD is shown in Figure \ref{f-mock2_psd}. Note that because this light curve is irregularly sampled, we do not compute a periodogram due to distortions caused by the sampling pattern.

\begin{figure}
	\includegraphics[scale=0.45, angle=00]{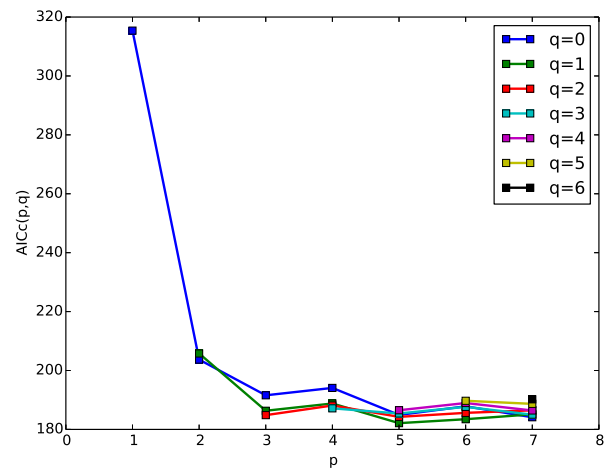}
	\caption{The AICc values from the simulated light curve shown in Figure \ref{f-mock2_lcurve} for CARMA($p,q$) models of order $p \leq 7, q < p$. The minimum AICc is achieved for a value of $p = 5, q = 1$, although the AICc curve is fairly flat for $p \geq 3$.
	\label{f-mock2_aicc}}
\end{figure}

We ran our MCMC sampler on the second light curve using the same configuration as for the first. The AICc values are shown in Figure \ref{f-mock2_aicc}. In this case, the $p = 5, q = 1$ model was chosen as having the best AICc. The $95\%$ probability bounds on the PSD based on the CARMA(5,1) model are also shown in Figure \ref{f-mock2_psd}. The CARMA(5,1) model is able to recover the PSD above the measurement noise level. The high-frequency oscillatory feature may be encompassed in the probability contours derived from a higher order CARMA process, however a `detection' of this feature would not be possible. In an actual analysis one would in general not have knowledge of the PSD below the measurement noise level, so we consider it best practice to use the simplest model that adequately describes the variability features above the measurement noise level.

Finally, as an illustration for applications where one may desire interpolated or forecasted values of a lightcurve, in Figure \ref{f-mock2_lcurve} we also show the interpolated and forecasted values of the simulated lightcurve, along with their $1\sigma$ uncertainties, based on the best-fit CARMA(5,1) model. These quantities provide a means of simulating realizations of the light curve at these time points, conditional on the measurement light curve, as described in \S~\ref{s-interpolation}.

\subsection{Non-Stationary Process Under Irregular Sampling}
\label{s-nonstationary_irregular}

In order to illustrate what a CARMA model fit would look like for a non-stationary process, we simulated a light curve that switches from one CARMA process to another. In reality the behavior of a CARMA model fit to a non-stationary process will depend on the nature of the non-stationarity, and this simple illustration is meant to provide some qualitative insight into how non-stationarity affects the inferred PSD. Moreover, we also note that non-stationarity could be modeled by allowing the CARMA process parameters to change with time.

\begin{figure}
	\includegraphics[scale=0.45,angle=00]{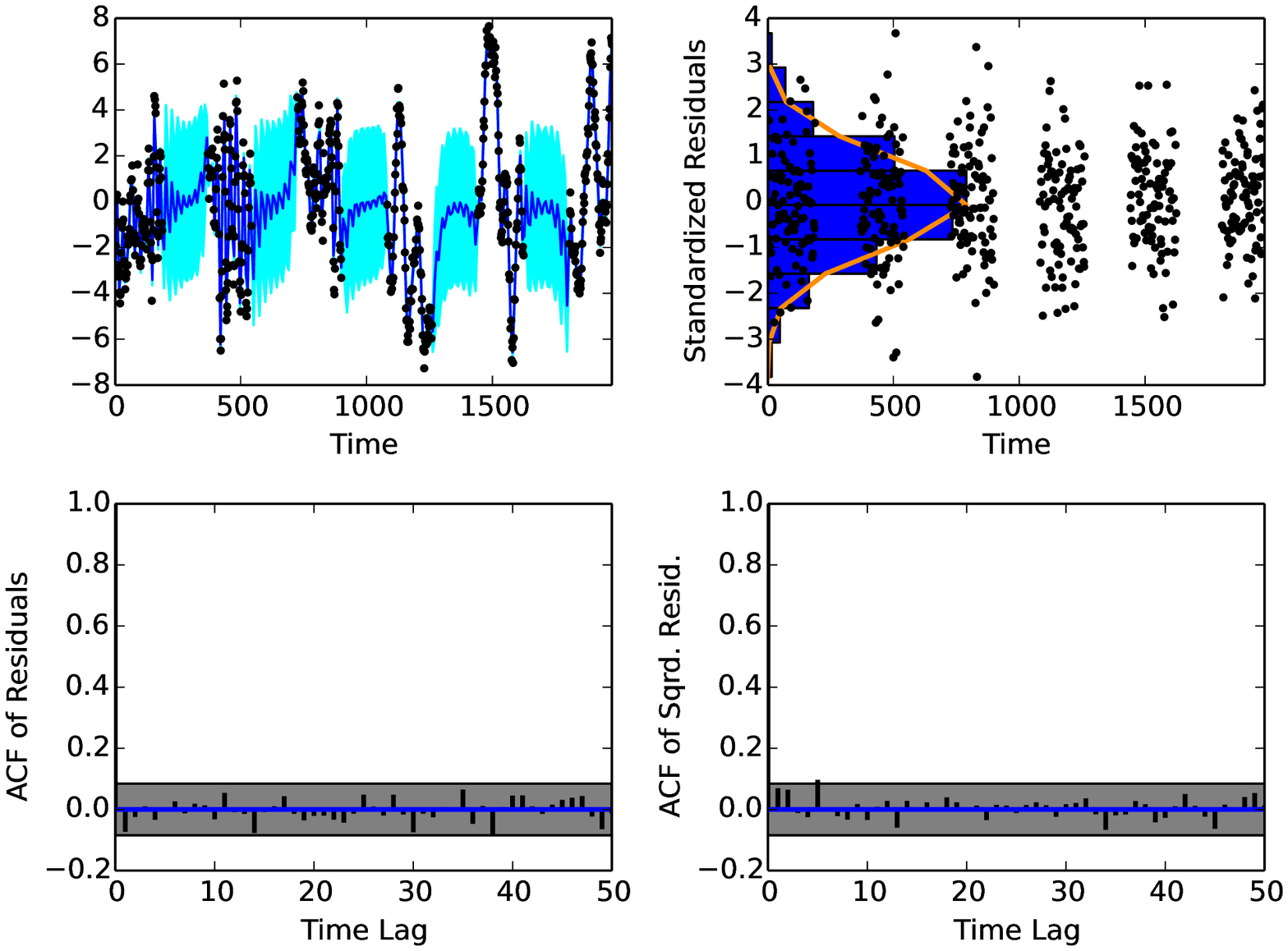}
	\caption{Clockwise from top left: a) Light curve for a simulated non-stationary lightcurve, as well as the interpolated values based on the best-fitting CARMA(5,2) process. Symbols are as in Figure \ref{f-mock2_lcurve}. b) Standardized residuals (data points) and their distribution (blue histogram), compared with the expected standard normal distribution (orange line). There is no evidence for a deviation from a Gaussian CARMA process for this light curve. c) and d) Autocorrelation functions of the standardized residuals (bottom left) and their square (bottom right), compared with the 95\% confidence region assuming a white noise process (shaded region). There is no evidence that the residuals deviate from a white noise sequence, suggesting that the CARMA model has adequately captured the correlation structure in the light curve.
	\label{f-nonstationary}}
\end{figure}

We constructed a non-stationary light curve by generating two separate light curves with the same sampling scheme as for the stationary irregularly sampled light curve above. For the first process we used the CARMA parameters from \S~\ref{s-stationary_irregular}, while for the second we used the parameters from \S~\ref{s-stationary_regular}. In addition, the variance of the latter CARMA process was set to be twice that of the former. We constructed a non-stationary light curve by setting the first half to the former process, and the second half to the latter process. The light curve is shown in Figure \ref{f-nonstationary}.

\begin{figure}
	\includegraphics[scale=0.45,angle=00]{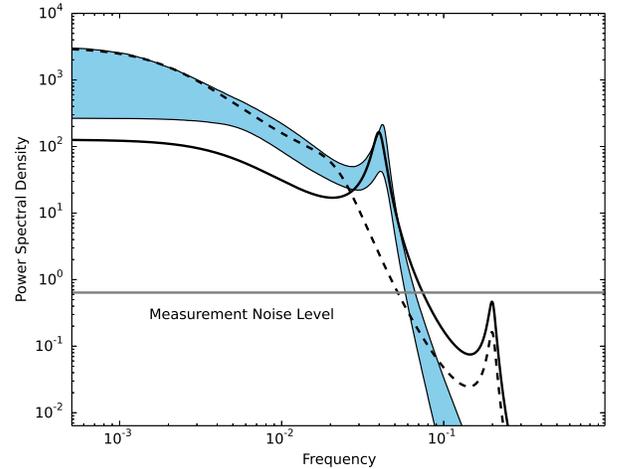}
	\caption{Power spectral density for the non-stationary light curve shown in Figure \ref{f-nonstationary}. The blue region contains $95\%$ of the posterior probability, the solid line shows the PSD for the first half of the light curve, and the dashed line shows the PSD for the second half of the light curve. The constraints on the PSD are given for a $p = 5, q = 2$ model. The inferred PSD assuming the stationary CARMA model is a blend of the two true PSDs, picking up both the strong QPO present in the first half of the light curve and the strong low-frequency broad-band noise in the second half of the light curve.
	 \label{f-nonstationary_psd}}
\end{figure}

Minimizing the AICc chose a CARMA(5,2) model. The fit quality is shown in Figure \ref{f-nonstationary} and the PSD is shown in Figure \ref{f-nonstationary_psd}. Based on distribution and ACF of the residuals there is no statistically significant evidence for a deviation from a single CARMA process for this light curve, suggesting that non-stationarity may be difficult to detected using these diagnostics, at least at this data quality. The inferred PSD is a blend of the two separate CARMA processes, picking up the dominant sources of variability. In particular, the PSD picks up the strong QPO present in the first half of the lightcurve, and large amount of broad-band variability power at the longest time scales present in the second half of the light curve. This is, in a sense, because these features have the strongest `signal-to-noise' ratio, as their variability amplitude and frequency distribution is most distinguished from the measurement noise and is constrained by the frequency range probed by the sampling pattern. From this we infer that the inferred PSD for a non-stationary light curve will be a weighted average of a time-varying PSD over the observing period of a light curve, where the weights are strongest for PSDs and frequencies that have the highest signal-to-noise. 

\section{Example Applications to Real Lightcurves}
\label{s-applications}

In this section we illustrate the application of CARMA models to a variety of astronomical variables, including an X-ray lightcurve of X-ray binaries, optical light curves of AGN, optical lightcurves of two variable stars. 

\subsection{X-ray Binary}
\label{s-xray_binaries}

\begin{figure}
	\includegraphics[scale=0.45,angle=00]{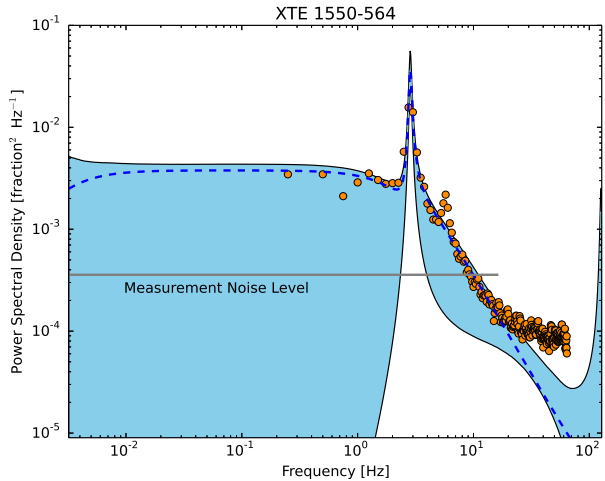}
	\caption{Power spectral density for the \emph{RXTE} light curve of the X-ray binary XTE 1550-564, with symbols the same as for Figure \ref{f-mock1_psd}. The constraints on the PSD are given by a $p = 5, q = 4$ model. The periodogram corresponds to the full light curve segment with $4 \times 10^4$ data points, while the CARMA constraints are obtained after randomly downsampling the light curve to 4000 data points. In addition, the length of the line marking the approximate measurement error noise level in the down sampled light curve also marks the frequency range probed by the down sampled light curve, with the upper limit corresponding to the average value of the Nyquist Frequency, $\langle 1 / 2 \Delta t \rangle$. We note that the measurement noise level of the down-sampled light curve is higher relative to the full light curve by the ratio of Nyquist frequencies, since the integrated power in the measurement noise is preserved. The CARMA model recovers the dominant QPO, but does not find evidence for the higher frequency weaker QPO. It likely misses the higher frequency QPO because it lies at the upper limit of the frequency range probed by the down sampled light curve.
	 \label{f-xte_psd}}
\end{figure}

We first apply our CARMA modeling to a Rossi X-ray Timing Explorer (\emph{RXTE}) light curve of the X-ray binary XTE 1550-664; the OBSID of this light curve is 30191-01-16. The data reduction for this light curve is described in \citet{Heil2011a}. This light curve was chosen because it is densely and regularly sampled every $1 / 128$ seconds, and because it has a complex and well-measured PSD with multiple QPOs. We analyze a $\approx 312$ sec segment of this light curve from $t = 848.02$ sec to $t = 1160.47$ sec. The full light curve has $4 \times 10^4$ data points, and its periodogram is shown in Figure \ref{f-xte_psd}. The PSD is flat on time scale longer than $\approx 1$ sec, and shows a strong QPO at $\approx 3$ Hz; there is a second weaker QPO at $\approx 6$ Hz. The flattening in the PSD at the highest frequencies is caused by the measurement noise.

Before applying our CARMA modeling we randomly downsampled the light curve to 4000 data points. This therefore provides an interesting test of the CARMA modeling to recover the PSD from an irregularly-sampled light curve from an astronomical source for which the PSD is effectively known. The mean time spacing of the down-sampled light curve is $\Delta t = 0.078$ sec, and the measurement noise contributes $\approx 8\%$ to the observed root-mean-square of the natural logarithm of the flux values (i.e., the ratio of measurement noise standard deviation to the observed RMS of the natural logarithm of the measured flux values is $\approx 0.08$). Because the X-ray binary light curves have a log-normal distribution \citep{Uttley2005a}, and because flux must be non-negative, we applied our CARMA modeling to the logarithms of the flux values.

The CARMA model that minimized the AICc had $p = 4, q = 3$. The existence of two QPOs in addition to a broad-band noise component in the periodogram of the full light curve implies at least $p = 5$, because the number of QPO features that can be modeled for a CARMA process of order $p$ is $\lfloor p/2 \rfloor$, where $\lfloor \cdot \rfloor$ is the floor function, and because a broad-band noise component (i.e., a zero-centered Lorentzian function in the PSD) always occurs for odd values of $p$. This therefore suggests that the AICc value does not find sufficient improvement to justify the inclusion of the weaker QPO feature. In order to assess the statistical significance of the feature we ran our MCMC sampler using values of $p = 5, q = 4$. The PSD from the CARMA(5,4) model is also shown in Figure \ref{f-xte_psd}. The dominant QPO is clearly recovered, with an estimated centroid of $2.84 \pm 0.01$ Hz and quality factor of $Q = 16.32 \pm 3.27$; the QPO quality factor is the ratio of the Lorentzian centroid frequency to its FWHM. However, the weaker QPO is missed suggesting that there is not evidence for it in the down-sampled light curve. This is likely due to the fact that its centroid is close to the average Nyquist frequency of the down sampled light curve, and thus falls at the edge of the frequency range probed. In addition, there is no evidence in the down sampled light curve for significant additional variability power on time scales longer than the QPO, much less evidence that it flattens to white noise.

\begin{figure*}
	\includegraphics[scale=0.31,angle=00]{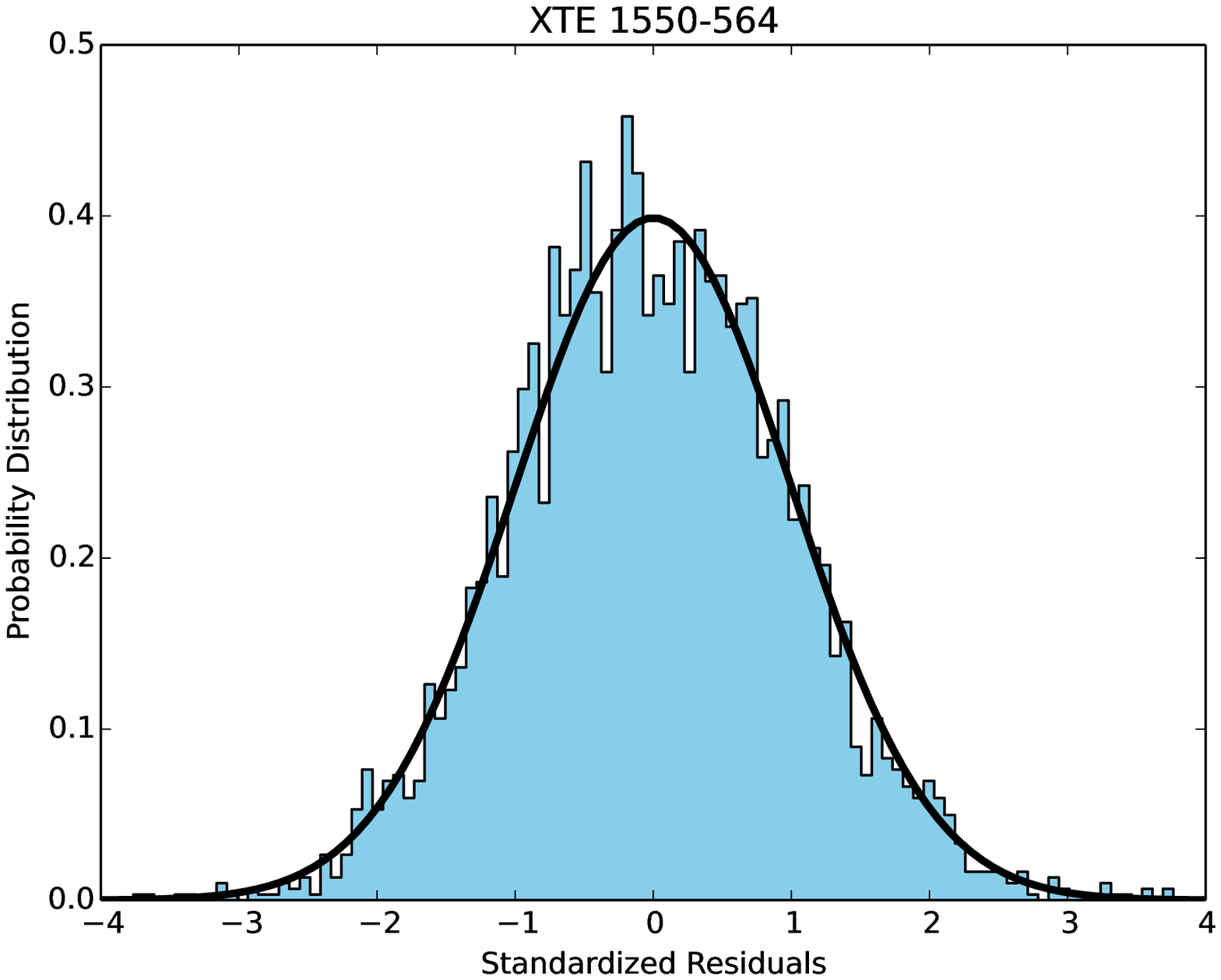}
	\includegraphics[scale=0.31,angle=00]{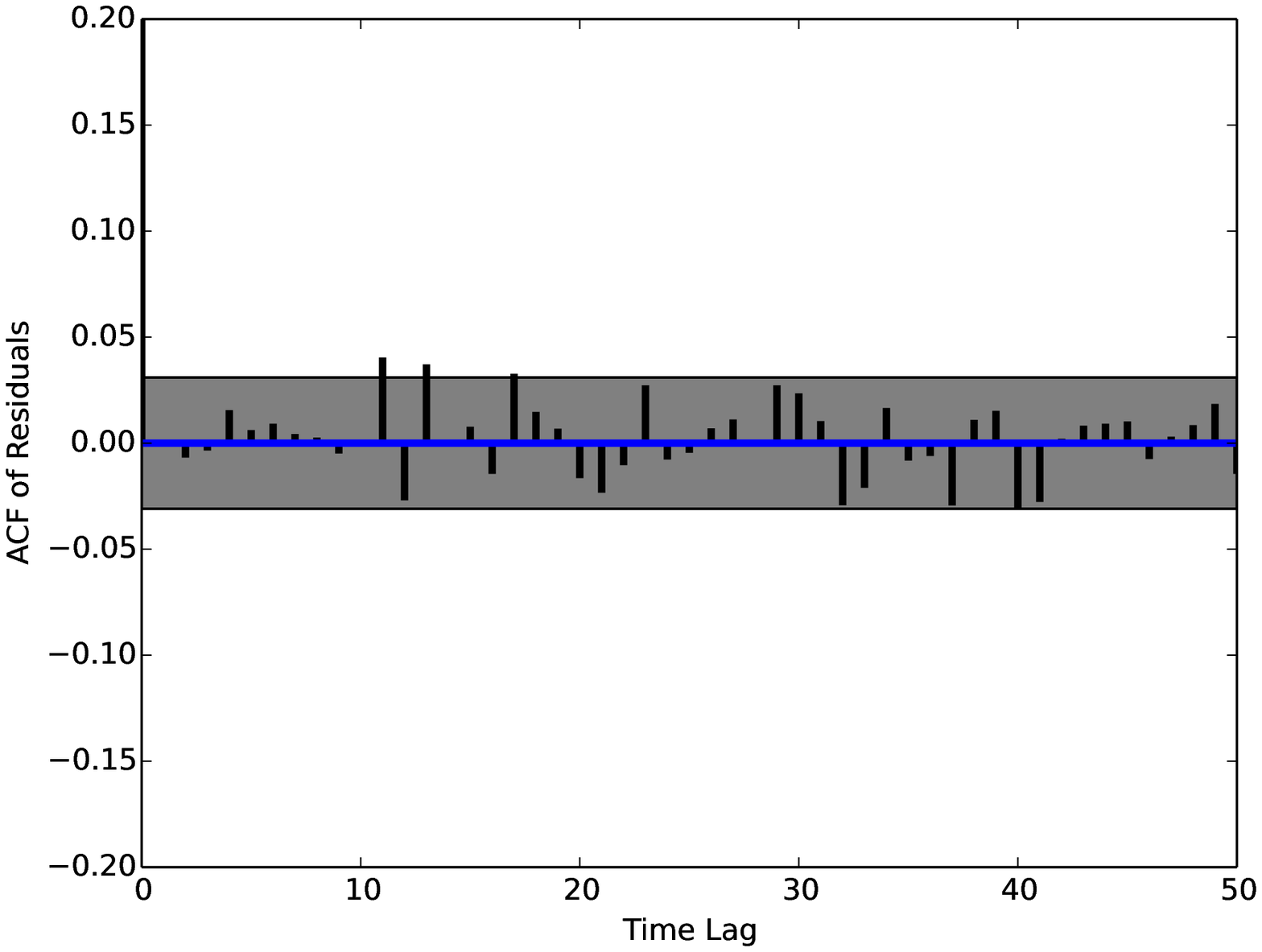}
	\includegraphics[scale=0.31,angle=00]{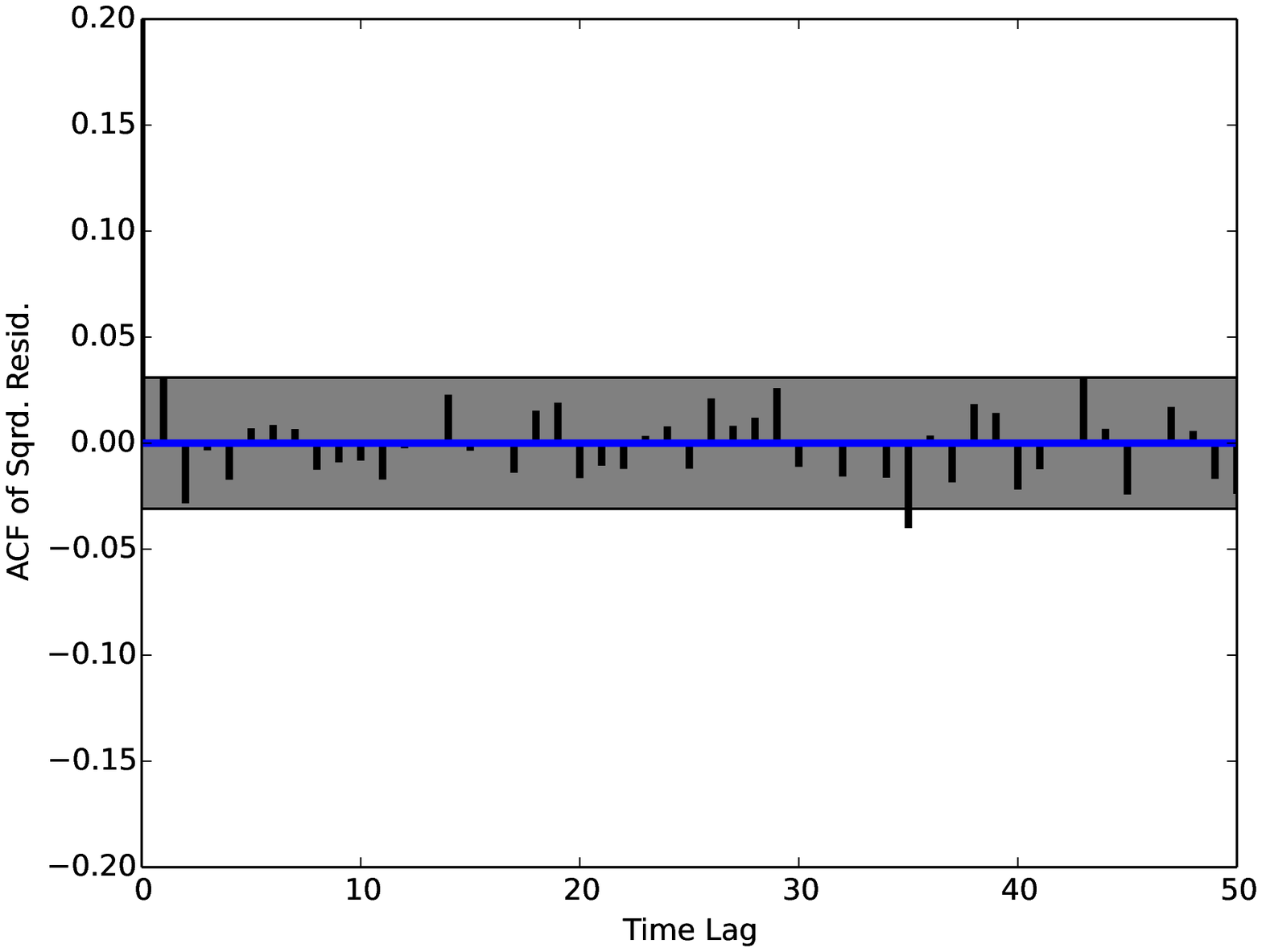}
	\caption{Histogram of the standardized residuals (left), autocorrelation function of the standardized residuals (center), and autocorrelation function of the square of the standardized residuals (right); all standardized residuals were calculated for a CARMA(5,4) model. The standardized residuals do not show any significant departures from a Gaussian distribution (solid black line), implying that the assumption of a Gaussian process is adequate. The $95\%$ confidence limit on the autocorrelation function of a white noise sequence is also shown in the middle and right plots, denoted by the grey region. The sequence of standardized residuals is consistent with a white noise process, implying that the CARMA(5,4) model has adequately captured the correlation structure in the light curve. Moreover, the sequence of the squares of the standarized residuals is also consistent with white noise, implying that there is not evidence for non-linear behavior in the light curve when using the logarithm of flux.
	 \label{f-xte_fit_quality}}
\end{figure*}

We assess the quality of the CARMA(5,4) model fit by inspection of the histogram of standardized residuals, and of the autocorrelation functions of the standardized residuals and their squares, all of which are plotted in Figure \ref{f-xte_fit_quality}. There is no evidence for a significant departure from the assumption of a Gaussian process, and the residuals are consistent with white noise, implying that the Gaussian CARMA(5,4) model adequately describes the fluctuations in this light curve.

\subsection{Active Galactic Nuclei}
\label{s-agn}

We applied our CARMA modeling to optical light curves of two AGN. The first light curve is from the \emph{Kepler} observatory, and the second is from the SDSS Stripe 82 data set.

\subsubsection{\emph{Kepler} Lightcurve}
\label{s-kepler}

\begin{figure}
	\includegraphics[scale=0.45,angle=00]{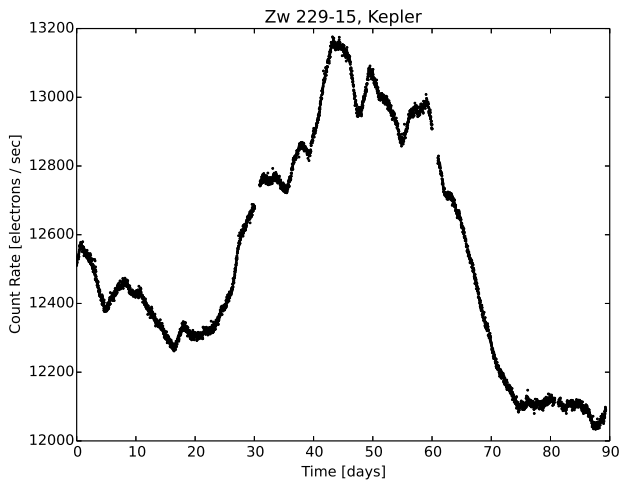}
	\caption{Optical light curve for the AGN Zw 229-15 from the \emph{Kepler} satellite in quarter Q9. 
	\label{f-kepler_lcurve}}
\end{figure}

\begin{figure}
	\includegraphics[scale=0.45,angle=00]{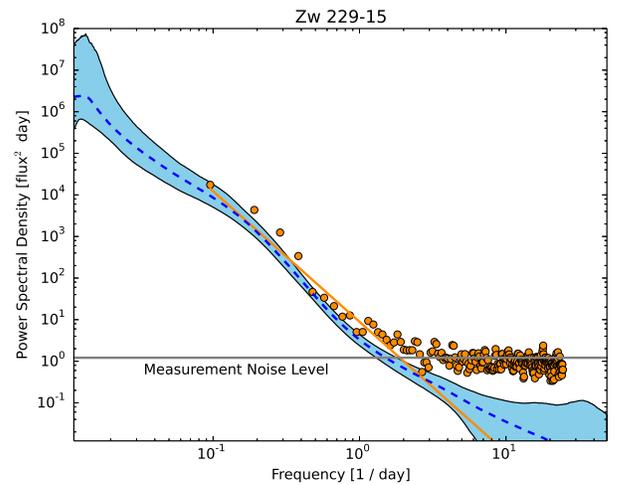}
	\caption{Power spectral density for the \emph{Kepler} light curve of Zw 229-15, with symbols the same as for Figure \ref{f-mock1_psd}. The constraints on the PSD are given by a $p = 6, q = 4$ model. The dark orange line shows a power-law PSD of the form $P(f) \propto 1 / f^{\alpha}$, where \citet{Mushotzky2011a} find a best-fit of $\alpha = 3.14$. The CARMA model PSD tracks the periodogram well, confirming that on time scales $\lesssim 10$ days the PSD for this source is steeper than expected from a CAR(1) model. The CARMA model also shows some evidence for the PSD flattening to $\sim 1 / f^2$ on time scales $\gtrsim 10$ days.
	 \label{f-kepler_psd}}
\end{figure}

The first optical light curve is from the \emph{Kepler} observatory for the local AGN Zw 229-15 ($z = 0.027879$) in quarter Q9. The Kepler light curves are the highest quality light curves for any AGN, with this one being almost regularly sampled every 30 minutes for approximately 3 months providing a total of 4375 data points. This therefore makes it a good test case for our CARMA modeling approach. In addition the amplitude of the measurement errors is only $\approx 1.5\%$ of the observed standard deviation in the light curve. This light curve was analyzed by \citet{Mushotzky2011a} who concluded that the PSD can be characterized as a power-law $P(f) \propto 1 / f^{3.14}$. The light curve is shown in Figure \ref{f-kepler_lcurve}, and the periodogram is shown in Figure \ref{f-kepler_psd}. When computing the periodogram we also employed the `end-matching' technique used by \citet{Mushotzky2011a}. Also shown in Figure \ref{f-kepler_psd} is a PSD of the form $P(f) \propto 1 / f^{3.14}$, as estimated by \citet{Mushotzky2011a}.

The AICc values were minimized for $p=6,q=4$. The standardized residuals using the best-fit model did not show any evidence for deviations from a Gaussian white noise process, implying that the CARMA(6,4) model is sufficient. The maximum-likelihood estimate and region containing $95\%$ of the posterior probability are both shown in Figure \ref{f-kepler_psd}. The PSD from the CARMA(6,4) model is very similar to the periodogram above the measurement noise level, and can be well-approximated as a power-law with the slope $\sim -3$ on time scales shorter than $\approx 1$ month, consistent with the analysis of \citet{Mushotzky2011a}. However, there is some evidence that the PSD flattens to $\sim 1 / f^2$ on time scales $\gtrsim 10$ days. If real and common among AGN, this may explain why a CAR(1) model has been successful in modeling optical AGN variability on these longer time scales.

\subsubsection{SDSS Stripe 82 Lightcurve}

\begin{figure}
	\includegraphics[scale=0.45,angle=00]{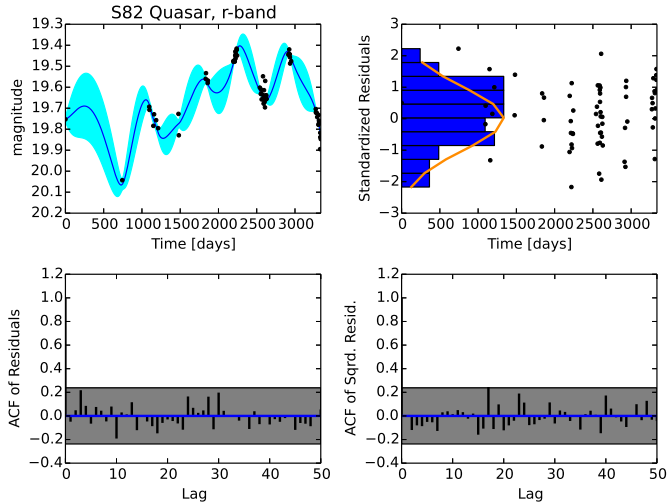}
	\caption{Clockwise from top left: a) $r$-band light curve for a quasar from the SDSS Stripe 82 survey, as well as the interpolated values based on the best-fitting CARMA(4,1) process. Symbols are as in Figure \ref{f-mock2_lcurve}. b) Standardized residuals (data points) and their distribution (blue histogram), compared with the expected standard normal distribution (orange line). There is no evidence for a deviation from a Gaussian CARMA process for this light curve. c) and d) Autocorrelation functions of the standardized residuals (bottom left) and their square (bottom right), compared with the 95\% confidence region assuming a white noise process (shaded region). There is no evidence that the residuals deviate from a white noise sequence, suggesting that the CARMA model has adequately captured the correlation structure in the light curve.
	\label{f-s82qso_lcurve}}
\end{figure}

An r-band AGN light curve from Stripe 82 is taken from the catalogue of \citet{MacLeod2012a} and is for the quasar with RA and Dec (J2000) 10:02:34.6, -00:59:19.5, and a redshift of $z = 1.5239$. This light curve has a time baseline of a little less than 10 years, but suffers from considerable irregular sampling.  The Stripe 82 light curve has 68 data points and the average amplitude of the measurement errors is $\approx 16\%$ of the observed standard deviation in the light curve. The light curve is shown in Figure \ref{f-s82qso_lcurve}.

\begin{figure}
	\includegraphics[scale=0.45,angle=00]{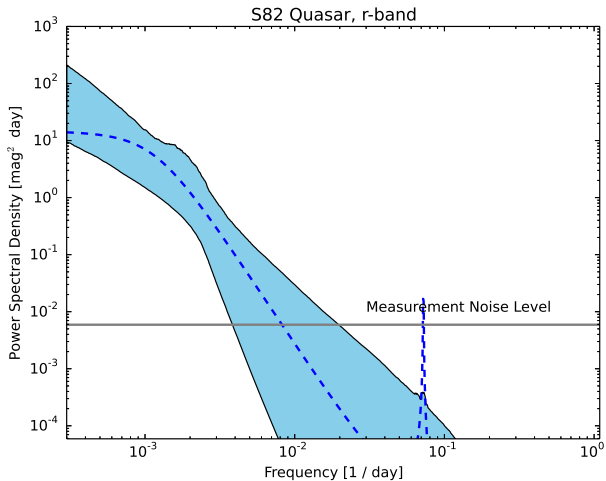}
	\caption{Power spectral density for the light curve of the Stripe 82 quasar, with symbols the same as for Figure \ref{f-mock1_psd}. The constraints on the PSD are given by a $p = 4, q = 1$ model. There is considerable uncertainty in the estimated PSD, and the PSD is consistent with a power law of the form $P(f) \propto 1 / f^{\alpha}, \alpha = 2$--$4$. In addition, there is some marginal evidence that the PSD flattens toward lower frequencies. Note that the high-frequency spike in the maximum-likelihood estimate is unlikely to be real, as it falls well below the measurement noise level.
		 \label{f-s82qso_psd}}
\end{figure}

The AICc values for this light curve were minimized at $p=4,q=1$. In Figure \ref{f-s82qso_lcurve} we compare the measured light curve with an interpolation based on the best-fit CARMA(4,1) model. Also shown in Figure \ref{f-s82qso_lcurve} are the standardized residuals and autocorrelation functions of the residuals and their square. There is no evidence for significant deviations from the assumption of a Gaussian process, and the residuals are consistent with white noise implying that a CARMA(4,1) process adequately describes the fluctuations of this light curve. The maximum-likelihood estimate of the PSD is shown in Figure \ref{f-s82qso_psd}, along with the region containing 95\% of the posterior probability. The PSD shows some evidence for steepening toward higher frequencies, although the uncertainties are large and a single power-law of the form $1 / f^{\alpha}, \alpha = 2$--$4$ is also consistent with the estimated PSD.

%The X-ray light curve is for the source MCG-6-30-15. The data are a combination of \emph{RXTE} and \emph{XMM} data, and there are a total of 3870 data points. The \emph{RXTE} data is from \citet{Sobolewska2009a}, and the \emph{XMM} data is from \citet{Kelly2011a}; the two data sets are combined as described in \citet{Kelly2011a}. The \emph{RXTE} data are irregularly sampled over a $\approx 10$ year period, and the \emph{XMM} data are regularly sampled every 48 seconds for $\sim 10^5$ sec. The amplitude of the measurement errors are $\approx 14\%$ and $\approx 36\%$ of the measured standard deviation in the light curve for the \emph{RXTE} and \emph{XMM} data, respectively. The X-ray PSD for MCG-6-30-15 was studied by \citet{McHardy2005a} and \citet{Kelly2011a}, both of whom found that the PSD is roughly characterized as having a broken power-law shape with $P(f) \propto 1 / f^{0.75}$ at frequencies $f \lesssim 7 \times 10^{-5}\ {\rm Hz}$, steepening to $P(f) \propto 1 / f^{\alpha}, \alpha \sim 2$ at higher frequencies.
%

\subsection{Variable Stars}
\label{s-variable_stars}

As an illustration we also applied our CARMA modeling to the optical variations of two variable stars. The first is a Long-period variable star, and the second is an RR-Lyrae star. Both show regular variations, with the RR-Lyrae stars variations being more regular and deterministic. They contrast with the X-ray binary and AGN light curves in that their emission does not come from an accretion disk, but instead is driven by pulsations within the stellar atmosphere.

\subsubsection{Long-Period Variable}
\label{s-lpvs}

\begin{figure}
	\includegraphics[scale=0.45,angle=00]{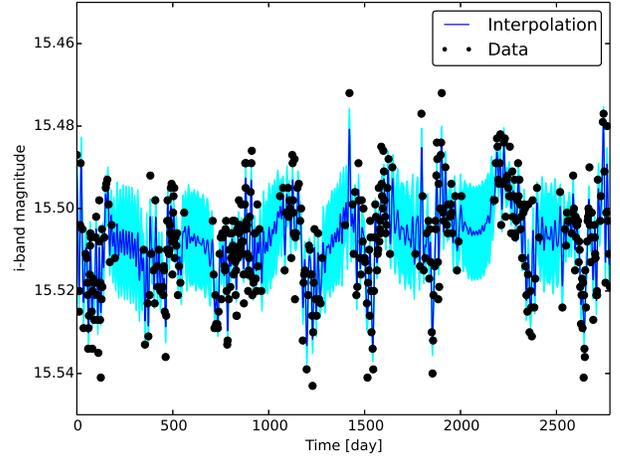}
	\caption{$i$-band light curve for a long-period variable star on the Red Giant Branch, from the OGLE-III survey. Also shown is the interpolated light curve and its uncertainty assuming a CARMA(6,0) model; the symbols are the same as in Figure \ref{f-mock2_lcurve}.}
	\label{f-lpv_lcurve}
\end{figure}

The first variable star light curve we applied our CARMA modeling to is the $i$-band light curve of a long-period variable on the red giant branch from the OGLE-III variable star catalogue \citep{Soszynski2011a}. The light curve is shown in Figure \ref{f-lpv_lcurve}. The RA and Dec of this source are 04:27:55.78, -70:24:59.4. The light curve for this source has 437 data points, spans $7.6$ years, and has a median time sampling of 3 days. This is a relatively low $S/N$ light curve relative to the intrinsic source variability, as the measurement errors make up $\approx 48\%$ of the observed standard deviation in the light curve. Part of our motivation for choosing this light curve as an example is because long-period variable stars are the main contaminant in samples of AGN selected using the CAR(1) parameters \citep{MacLeod2011a}, and hopefully the two sources become distinguishable using higher order CARMA models.

\begin{figure}
	\includegraphics[scale=0.45,angle=00]{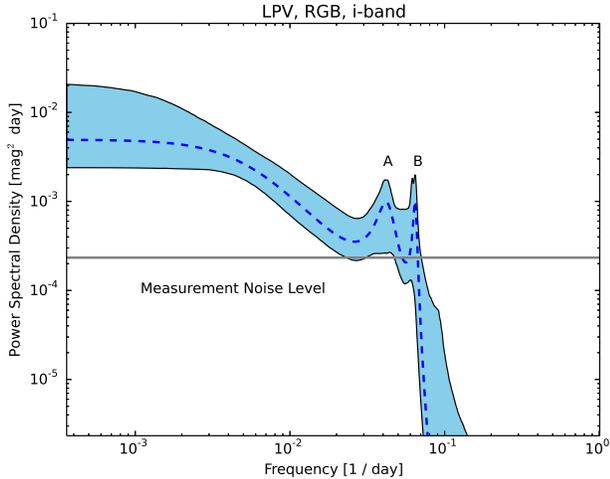}
	\caption{Power spectral density for the light curve of the long-period variable from the OGLE-III survey assuming a CARMA(6,0) model; symbols are the same as in Figure \ref{f-mock2_lcurve}. The power spectrum is flat on the longest time scales, implying uncorrelated variations on time scales $\gtrsim 187$ days. On shorter time scales the PSD flatten to $\sim 1 / f$. In addition, there are two pulsation modes with quasi-periods corresponding to $\approx 16$ (labeled "B") and $\approx 25$ days (labeled "A"), respectively. However, the lower-frequency QPO feature is only has a posterior probability of $\sim 63\%$.
		 \label{f-lpv_psd}}
\end{figure}

The AICc for this light curve was minimized at $p = 6, q = 0$. There was no significant evidence for deviations from the CARMA(6,0) model for this light curve, as the residuals were consistent with Gaussian white noise. In Figure \ref{f-lpv_lcurve} we also show the light curve interpolated from the best-fit CARMA(6,0). The estimated PSD is shown in Figure \ref{f-lpv_psd}. The PSD on the longest time scales is flat, implying uncorrelated variability, and steepens to $\sim 1 / f$ at a characteristic frequency $\omega = 1 / \tau$. We estimate $\tau = 187$ days with a $95\%$ credibility interval of $15.5 < \tau < 320$ days. The higher-frequency pulsation mode, labeled "B", corresponds to a quasi-periodic oscillation with a time-scale of 15.9 days (95\% credibility interval of $(11.1, 19.5)$ days). The existence of QPO "B" is highly significant, as it is present in all of the MCMC samples. QPO B has a posterior median quality factor of $Q = 6.85$ with a 95\% credibility interval of $1.5 < Q < 42$. The lower-frequency pulsation mode, labeled as "A", was not present in $\approx 37\%$ of the MCMC samples suggesting that it has a posterior probability of only $63\%$ and is therefore not statistically signifiant. The quasi-period found from the CARMA model for QPO B is shorter than the period of 22.46 days quoted in the OGLE-III catalogue \citep{Soszynski2011a}, although the A pulsation mode, if real, has a period of $\approx 25$ days.

\subsubsection{RR-Lyrae}
\label{s-rrlyrae}

\begin{figure}
	\includegraphics[scale=0.45,angle=00]{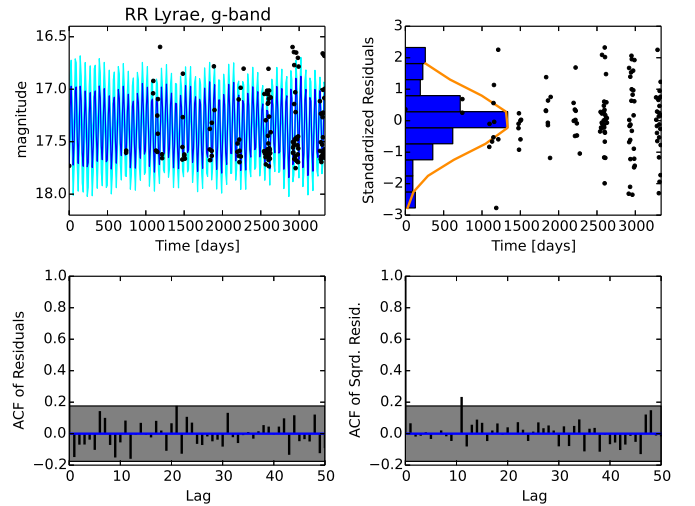}
	\caption{Clockwise from top left: a) $g$-band light curve for an RR-Lyrae star from the SDSS Stripe 82 survey, as well as the interpolated values based on the best-fitting CARMA(7,0) process. Symbols are as in Figure \ref{f-mock2_lcurve}. b) Standardized residuals (data points) and their distribution (blue histogram), compared with the expected standard normal distribution (orange line). The distribution of the residuals is considerably narrower than the standard normal, suggesting that the assumption of a Gaussian process is not appropriate for this lightcurve. c) and d) Autocorrelation functions of the standardized residuals (bottom left) and their square (bottom right), compared with the 95\% confidence region assuming a white noise process (shaded region). There is no evidence that the residuals deviate from a white noise sequence, suggesting that the CARMA model has captured the correlation structure in the light curve.
	\label{f-rrly_lcurve}}
\end{figure}

We also applied our CARMA modeling to the $g$-band light curve of an RR-Lyrae from the Stripe 82 catalogue of \citep{Sesar2010a}. The RA and Dec (J2000) of this source are 06:41:29.48, -00:00:01.68. There are 128 epochs in this light curve over $\sim 9$ years with a median time spacing of 2 days. The measurement errors are $\approx 3.5\%$ of the observed standard deviation in the light curve. RR-Lyrae sources show regular non-sinusoidal periodic fluctuations, and thus a stochastic process such as the CARMA model may not provide the best representation of their light curves. We include this application as an example of how the CARMA modeling performs for a periodic source. The light curve for this source is shown in Figure \ref{f-rrly_lcurve}.

A CARMA(7,0) model was found to minimize the AICc. The interpolated light curve based on the best-fit CARMA model is also shown in Figure \ref{f-rrly_lcurve}, as well as the standardized residuals and their autocorrelation functions. The autocorrelation functions do not show any significant deviations from white noise, suggesting that the CARMA(7,0) model has captured the correlation structure in the light curve within the limits of the data quality. However, the histogram of the residuals is more narrow than a normal distribution, suggesting deviations from the assumption of a Gaussian stochastic process. This is not surprising, as RR-Lyrae exhibit regular periodic variations and thus a CARMA model is unlikely to be the best choice. We note that the consistency of the residuals with a white noise sequence implies that it is not necessary for the residuals to be normally distributed in order for the CARMA model to capture much of the correlation structure in a light curve.  In addition, the significant deviation in the residuals from a normal distribution may provide a way of using the CARMA process parameters to discriminate between periodic and aperiodic variables in time-domain surveys.

\begin{figure}
	\includegraphics[scale=0.45,angle=00]{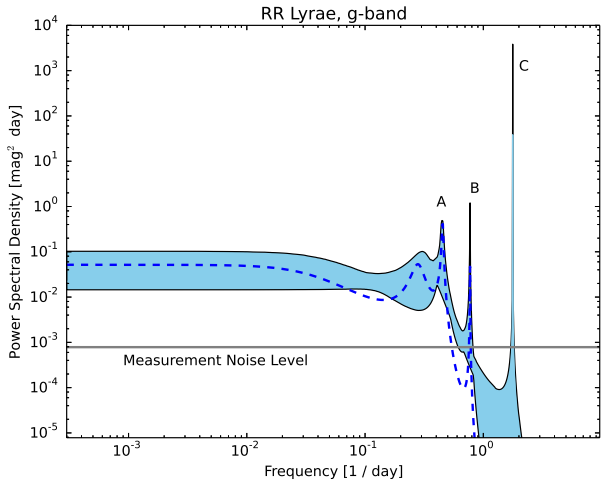}
	\caption{Power spectral density for the $g$-band light curve of an RR-Lyrae star from the SDSS Stripe 82, assuming a CARMA(7,0) model; symbols are the same as in Figure \ref{f-mock2_lcurve}. The power spectrum is flat on the longest time scales, implying uncorrelated variations on time scales $\gtrsim 10$ days. The three pulsation modes are labeled "A", "B", and "C", and are in increasing order in terms of their centroid frequency. There are two statistically-significant pulsation modes in the PSD, with the mode A being a little broader and corresponding to a period of $\approx 2.5$ days. While it appears that there are two significant higher-frequency pulsation modes (B and C), only one of the two is present at any time in the MCMC samples. The high-frequency pulsation occurs at either a period of $\approx 1.3$ days (mode B) or $\approx 0.56$ days (mode C), with the shorter period being $\sim 3$ times more likely.
		 \label{f-rrly_psd}}
\end{figure}

The inferred PSD for this light curve is shown in Figure \ref{f-rrly_psd}. The PSD is dominated by two narrow pulsation modes (labeled as "B" and "C"), plus a broader mode at lower frequency (labeled as "A"), and is flat on time scales $\gtrsim 10$ days. Modes B and C are mutually exclusive, in the sense that if mode B is present in an MCMC sample, mode C is not.  We used a clustering algorithm on the PSD Lorentzian centers and widths in order to identify which MCMC samples correspond to each pulsation mode, as the labeling used by the MCMC sampler for each Lorentzian does not uniquely map to a quasi-periodic feature in the PSD. Mode C corresponds to a period of 0.56 days and is observed in $\approx 75\%$ of the MCMC samples. This mode corresponds very closely to the catalogue period of 0.564 \citep{Sesar2010a}, found using the super smoother algorithm \citep{Reimann1994a}.  Mode B corresponds to a period of 1.30 days and is present in $\approx 23\%$ of the MCMC samples. Because the existence of the two modes is mutually exclusive, the fact that one of the two is present is statistically significant at $98\%$ probability, with mode C being the more likely of the two. Mode A is also statistically significant, being present in $99.986\%$ of the MCMC samples, and corresponds to a period of 2.49 days with a $95\%$ credibility interval of $(2.18, 3.18)$.

We note that getting our algorithm to converge was particularly problematic for this light curve, as we would often get quantitatively different results for different runs of the algorithm. Moreover, the maximum-likelihood estimate is not contained within the $95\%$ probability bounds found from the MCMC sampler. This is likely because the MCMC sampler has found a better solution due to the fact that it runs for many more iterations, while the optimizer used to compute the maximum-likelihood estimate only found suboptimal modes. These facts, in addition with the fact that the light curve is inconsistent with a Gaussian process, imply that the likelihood function is noisy and has many modes. This implies that optimizers and MCMC samplers that are robust against multi-modality and complicated likelihood spaces may be necessary in order to get reliable results from CARMA modeling of light curves that exhibit regularly periodic and non-sinusoidal variations.

\section{Discussion}
\label{s-discussion}

In this work we have introduced continuous-time autoregressive moving average processes as flexible models for stochastic and quasi-periodic light curves. These models account for irregular sampling and measurement errors, making them applicable to a wide variety of light curves. Moreover, they are flexible, as their PSD can be described as a sum of Lorentzian functions. The primary purposes of this modeling approach are 1) to provide a flexible way to estimate power spectra for astronomical light curves, and 2) to provide variability features for light curves that may be used in variability selection techniques and potentially in the identification of new classes of variables. Because one can compute the likelihood function for a light curve under a CARMA model they have the advantage that they are statistically efficient and rigorous, as all of the information in the light curve is used to estimate the variability parameters and inference is based on the well-developed statistical theory of maximum-likelihood or Bayesian inference. Moreover, calculation of the likelihood function is computationally efficient, scaling linearly with the number of data points in a light curve. This last point makes their application to massive time domain data sets particularly attractive.

Previous work has also expanded upon the CAR(1) model to introduce additional flexibility. \citet{Kelly2011a} developed a mixture of CAR(1) processes as a model for X-ray variability of AGN. In this model, the light curve is expressed as a weighted sum of independent CAR(1) processes with different characteristics time scales that are constrained to lie on a regular logarithmic grid. The free parameters for this model are the maximum and minimum characteristic time scales of the grid, the mean and variance of the lightcurve, and the weights. Comparison with Equation (\ref{eq-carma_autocovar} that the mixture of $p$ CAR(1) processes model of \citet{Kelly2011a} is a special case of a CARMA(p,q) process where the roots of Equation (\ref{eq-characteristic}) are constrained to be real. \citet{Kelly2011a} also showed that a mixture of CAR(1) processes can closely approximate a broken-power law model for the PSD. In this case, the maximum and minimum characteristic time scales correspond to the low and high frequency breaks, respectively, and the the sequence of weights can be calculated as a function of the slope of the power-law between the low and high frequency breaks, so long as the slope of the PSD is constrained to the range $[-2, 0]$. These constraints reduce the number of free parameters to five for this model, which can be a computational advantage. In addition, \citet{Kelly2011a} also showed that the solution to the stochastic linear diffusion equation is a mixture of CAR(1) processes, providing a physical interpretation of the variability model.

\citet{Andrae2013a} investigated extensions to the Gaussian CAR(1) process as a model for quasar variability. The processes investigated by \citet{Andrae2013a} included models with more flexible PSDs, such as a CAR(2) process and a CARMA(1,1) process. The former process provides the ability to capture quasi-periodic variations, while the latter process introduces additional smoothing of the stochastic driving noise. Both processes are special cases of the general class of CARMA processes we discuss here; however, we note that the CARMA(1,1) model is not stationary. \citet{Andrae2013a} also investigated CAR(1) models after relaxing the assumptions of a linear Gaussian process. In particular, they also investigated non-Gaussian CAR(1) models and non-linear processes where the light curve variance also stochastically varies in time. These models, while not as flexible in modeling the PSD as a CARMA process, provide valuable alternatives to linear Gaussian models and may provide a better description of the variability of some light curves. However, in the case of quasar light curves from Stripe 82 \citet{Andrae2013a} concluded that the linear Gaussian CAR(1) process provided the best model for most of the quasars, based on a Bayesian model comparison.

While CARMA models are, in theory, computationally efficient, the likelihood space can be complex and exhibit multiple modes which can be problematic for numerical optimizers and samplers. This is especially true for higher order models, and seems to affect higher order $q$ models more strongly than higher order $p$ models. The likelihood space may also be very complex for light curves that have regular nearly deterministic variations, such as RR-Lyrae stars. We consider the difficulty in optimizing or sampling from a multimodal complex posterior to be the primary short-coming of these models at this time, and dampens their computational efficiency. Thus, researchers who utilize them must be careful to check that the primary posterior mode has been found, and that the dominant posterior modes have been sampled from. We deal with this in our MCMC sampler using a parallel tempering algorithm, which we have used with success. However, even this can fail to adequately sample the posterior if an insufficient number of chains are run, or if the algorithm is not run for a sufficiently long period of time. Moreover, our maximum-likelihood estimation is rather simple, as we simply use 100 random starts and optimize by finding a local mode using a greedy gradient-based algorithm. Future work should focus on improving the effectiveness and efficiency of the maximum-likelihood estimate and MCMC algorithm.

In order to illustrate the applicability of CARMA models for a variety of astronomical sources, as well as to provide a guide for interpreting their results, we applied these models to light curves for an X-ray binary, two AGN, a long-period variable star, and a RR-Lyrae star. In general we found that the CARMA models provide a good description of these light curves, suggesting that they can be applied to a broad range of astronomical sources that exhibit stochastic or quasi-periodic variations. The only exception was the RR-Lyrae star light curve. This is to be expected, as RR-Lyrae light curves exhibit regular non-sinusoidal variations and thus are unlikely to have a strong stochastic component. However, in spite of this the CARMA models still identified the period quoted by the catalogue that this light curve was taken from in $\approx 75\%$ of the MCMC samples. Moreover, the deviation in the residuals from a normal distribution for the RR-Lyrae star implies that the distribution of the residuals from a CARMA fit may provide an effective means of discriminating between different types of variables, even for those for which a CARMA model is not optimal. These results suggests that there may be value in using variability features derived from CARMA parameters even for non-stochastic light curves. 

Further improvements to the CARMA modeling approach can be obtained by including a deterministic component, such as a periodic function, for modeling periodic light curves, such as those from RR-Lyrae. In this case the residuals from fitting a deterministic function are modeled as following a CARMA process. In fact, this is the motivation behind the Periodic Autoregressive Moving Average models \citep[PARMA, e.g.,][]{Anderson2013a}, which allow for periodic variations in the mean and autocovariance function of a time series. In addition, it is possible to define multivariate CARMA models through a vector- and matrix-valued extension to Equation (\ref{eq-carma}) \citep[e.g.,][]{Marquardt2007a,Schlemm2012a}. Multivariate CARMA models hold considerable potential for characterizing the full multi-passband variability information obtained by time-domain surveys, and will be the subject of future work. 

In summary, CARMA models provide an important addition to the astronomer's statistical toolbox in the era of massive time-domain surveys, and have the potential to play an important role in the analysis of variability as a probe of astrophysics, as well as in the use of variability as a means of identifying classes of astronomical sources. 

\acknowledgements

We would like to thank Tommaso Treu for helpful discussions and comments on our manuscript, Lucy Heil for graciously provided the light curve for XTE 1550-564 and helpful comments on our manuscript, Simon Vaughan for helpful comments on our manuscript, and an anonymous referee for helping improve our discussion of CARMA models. BK acknowledges support from the Southern California Center for Galaxy Evolution, a multi-campus research program funded by the University of California Office of Research. A.C.B. acknowledges funding from NASA Origins grant NNX09AB32G. M.S. was partially supported by the EU FP7 grant No. 312789, and by the Polish NCN grant No. 2011/03/B/ST9/03459. 

This paper includes data collected by the Kepler mission. Funding for the Kepler mission is provided by the NASA Science Mission directorate. Funding for the SDSS and SDSS-II has been provided by the Alfred P. Sloan Foundation, the Participating Institutions, the National Science Foundation, the U.S. Department of Energy, the National Aeronautics and Space Administration, the Japanese Monbukagakusho, the Max Planck Society, and the Higher Education Funding Council for England. The SDSS Web Site is http://www.sdss.org/.

The SDSS is managed by the Astrophysical Research Consortium for the Participating Institutions. The Participating Institutions are the American Museum of Natural History, Astrophysical Institute Potsdam, University of Basel, University of Cambridge, Case Western Reserve University, University of Chicago, Drexel University, Fermilab, the Institute for Advanced Study, the Japan Participation Group, Johns Hopkins University, the Joint Institute for Nuclear Astrophysics, the Kavli Institute for Particle Astrophysics and Cosmology, the Korean Scientist Group, the Chinese Academy of Sciences (LAMOST), Los Alamos National Laboratory, the Max-Planck-Institute for Astronomy (MPIA), the Max-Planck-Institute for Astrophysics (MPA), New Mexico State University, Ohio State University, University of Pittsburgh, University of Portsmouth, Princeton University, the United States Naval Observatory, and the University of Washington.

Our software made use of the Armadillo C++ linear algebra library \citep{Sanderson2010a}.

%\bibliography{bkelly_bib}

\begin{thebibliography}{}
\expandafter\ifx\csname natexlab\endcsname\relax\def\natexlab#1{#1}\fi

\bibitem[{{Akaike}(1973)}]{Akaike1973a}
{Akaike}, H. 1973, in Proceedings of the Second International Symposium on
  Information Theory, ed. B.~Petrov \& F.~Csaki (Budapest: Akademiai Kiado),
  267--281

\bibitem[{Anderson {et~al.}(2013)Anderson, Meerschaert, \&
  Zhang}]{Anderson2013a}
Anderson, P.~L., Meerschaert, M.~M., \& Zhang, K. 2013, Journal of Time Series
  Analysis, 34, 187

\bibitem[{{Andrae} {et~al.}(2013){Andrae}, {Kim}, \&
  {Bailer-Jones}}]{Andrae2013a}
{Andrae}, R., {Kim}, D.-W., \& {Bailer-Jones}, C.~A.~L. 2013, \aap, 554, A137

\bibitem[{Belcher {et~al.}(1994)Belcher, Hampton, \& Wilson}]{Belcher1994a}
Belcher, J., Hampton, J.~S., \& Wilson, G.~T. 1994, Journal of the Royal
  Statistical Society. Series B (Methodological), 56, pp. 141

\bibitem[{{Belloni}(2010)}]{Belloni2010a}
{Belloni}, T.~M. 2010, in Lecture Notes in Physics, Berlin Springer Verlag,
  Vol. 794, Lecture Notes in Physics, Berlin Springer Verlag, ed. T.~{Belloni},
  53

\bibitem[{{Brewer} {et~al.}(2011){Brewer}, {Treu}, {Pancoast}, {Barth},
  {Bennert}, {Bentz}, {Filippenko}, {Greene}, {Malkan}, \& {Woo}}]{Brewer2011a}
{Brewer}, B.~J., {Treu}, T., {Pancoast}, A., {et~al.} 2011, \apjl, 733, L33

\bibitem[{{Brockwell} \& {Davis}(2002)}]{Brockwell2002a}
{Brockwell}, P., \& {Davis}, R. 2002, Introduction to Time Series and
  Forecasting, 2nd edn. (New York, NY: Springer US)

\bibitem[{Broersen \& Bos(2006)}]{Broersen2006a}
Broersen, P. M.~T., \& Bos, R. 2006, Instrumentation and Measurement, IEEE
  Transactions on, 55, 1124

\bibitem[{{Butler} \& {Bloom}(2011)}]{Butler2011a}
{Butler}, N.~R., \& {Bloom}, J.~S. 2011, \aj, 141, 93

\bibitem[{{Choi} {et~al.}(2013){Choi}, {Gibson}, {Becker}, {Ivezi{\'c}},
  {Connolly}, {MacLeod}, {Ruan}, \& {Anderson}}]{Choi2013a}
{Choi}, Y., {Gibson}, R.~R., {Becker}, A.~C., {et~al.} 2013, ArXiv e-prints,
  arXiv:1312.4957

\bibitem[{{Done} {et~al.}(2007){Done}, {Gierli{\'n}ski}, \&
  {Kubota}}]{Done2007a}
{Done}, C., {Gierli{\'n}ski}, M., \& {Kubota}, A. 2007, \aapr, 15, 1

\bibitem[{{Done} {et~al.}(1992){Done}, {Madejski}, {Mushotzky}, {Turner},
  {Koyama}, \& {Kunieda}}]{Done1992a}
{Done}, C., {Madejski}, G.~M., {Mushotzky}, R.~F., {et~al.} 1992, \apj, 400,
  138

\bibitem[{{Drake} {et~al.}(2009){Drake}, {Djorgovski}, {Mahabal}, {Beshore},
  {Larson}, {Graham}, {Williams}, {Christensen}, {Catelan}, {Boattini},
  {Gibbs}, {Hill}, \& {Kowalski}}]{Drake2009a}
{Drake}, A.~J., {Djorgovski}, S.~G., {Mahabal}, A., {et~al.} 2009, \apj, 696,
  870

\bibitem[{{Emmanoulopoulos} {et~al.}(2013){Emmanoulopoulos}, {McHardy}, \&
  {Papadakis}}]{Emmanoulopoulos2013a}
{Emmanoulopoulos}, D., {McHardy}, I.~M., \& {Papadakis}, I.~E. 2013, \mnras,
  433, 907

\bibitem[{{Emmanoulopoulos} {et~al.}(2010){Emmanoulopoulos}, {McHardy}, \&
  {Uttley}}]{Emmanoulopoulos2010a}
{Emmanoulopoulos}, D., {McHardy}, I.~M., \& {Uttley}, P. 2010, \mnras, 404, 931

\bibitem[{{Frieman} {et~al.}(2008){Frieman}, {Bassett}, {Becker}, {Choi},
  {Cinabro}, {DeJongh}, {Depoy}, {Dilday}, {Doi}, {Garnavich}, {Hogan},
  {Holtzman}, {Im}, {Jha}, {Kessler}, {Konishi}, {Lampeitl}, {Marriner},
  {Marshall}, {McGinnis}, {Miknaitis}, {Nichol}, {Prieto}, {Riess}, {Richmond},
  {Romani}, {Sako}, {Schneider}, {Smith}, {Takanashi}, {Tokita}, {van der
  Heyden}, {Yasuda}, {Zheng}, {Adelman-McCarthy}, {Annis}, {Assef},
  {Barentine}, {Bender}, {Blandford}, {Boroski}, {Bremer}, {Brewington},
  {Collins}, {Crotts}, {Dembicky}, {Eastman}, {Edge}, {Edmondson}, {Elson},
  {Eyler}, {Filippenko}, {Foley}, {Frank}, {Goobar}, {Gueth}, {Gunn},
  {Harvanek}, {Hopp}, {Ihara}, {Ivezi{\'c}}, {Kahn}, {Kaplan}, {Kent},
  {Ketzeback}, {Kleinman}, {Kollatschny}, {Kron}, {Krzesi{\'n}ski}, {Lamenti},
  {Leloudas}, {Lin}, {Long}, {Lucey}, {Lupton}, {Malanushenko}, {Malanushenko},
  {McMillan}, {Mendez}, {Morgan}, {Morokuma}, {Nitta}, {Ostman}, {Pan},
  {Rockosi}, {Romer}, {Ruiz-Lapuente}, {Saurage}, {Schlesinger}, {Snedden},
  {Sollerman}, {Stoughton}, {Stritzinger}, {Subba Rao}, {Tucker}, {Vaisanen},
  {Watson}, {Watters}, {Wheeler}, {Yanny}, \& {York}}]{Frieman2008a}
{Frieman}, J.~A., {Bassett}, B., {Becker}, A., {et~al.} 2008, \aj, 135, 338

\bibitem[{Gardiner(2004)}]{Gardiner2004a}
Gardiner, C.~W. 2004, Handbook of stochastic methods for physics, chemistry and
  the natural sciences, 3rd edn., Springer Series in Synergetics (Berlin:
  Springer-Verlag), xviii+415

\bibitem[{{Graham} {et~al.}(2014){Graham}, {Djorgovski}, {Drake}, {Mahabal},
  {Chang}, {Stern}, {Donalek}, \& {Glikman}}]{Graham2014a}
{Graham}, M.~J., {Djorgovski}, S.~G., {Drake}, A.~J., {et~al.} 2014, ArXiv
  e-prints, arXiv:1401.1785

\bibitem[{{Heil} {et~al.}(2011){Heil}, {Vaughan}, \& {Uttley}}]{Heil2011a}
{Heil}, L.~M., {Vaughan}, S., \& {Uttley}, P. 2011, \mnras, 411, L66

\bibitem[{{Horne} \& {Baliunas}(1986)}]{Horne1986a}
{Horne}, J.~H., \& {Baliunas}, S.~L. 1986, \apj, 302, 757

\bibitem[{{Horne} {et~al.}(1991){Horne}, {Welsh}, \& {Peterson}}]{Horne1991a}
{Horne}, K., {Welsh}, W.~F., \& {Peterson}, B.~M. 1991, \apjl, 367, L5

\bibitem[{{Hurvich} \& {Tsai}(1989)}]{Hurvich1989a}
{Hurvich}, C.~M., \& {Tsai}, C.-L. 1989, Biometrika, 76, 297

\bibitem[{{Ivezic} {et~al.}(2008){Ivezic}, {Tyson}, {Acosta}, {Allsman},
  {Anderson}, {Andrew}, {Angel}, {Axelrod}, {Barr}, {Becker}, {Becla},
  {Beldica}, {Blandford}, {Bloom}, {Borne}, {Brandt}, {Brown}, {Bullock},
  {Burke}, {Chandrasekharan}, {Chesley}, {Claver}, {Connolly}, {Cook},
  {Cooray}, {Covey}, {Cribbs}, {Cutri}, {Daues}, {Delgado}, {Ferguson},
  {Gawiser}, {Geary}, {Gee}, {Geha}, {Gibson}, {Gilmore}, {Gressler}, {Hogan},
  {Huffer}, {Jacoby}, {Jain}, {Jernigan}, {Jones}, {Juric}, {Kahn}, {Kalirai},
  {Kantor}, {Kessler}, {Kirkby}, {Knox}, {Krabbendam}, {Krughoff}, {Kulkarni},
  {Lambert}, {Levine}, {Liang}, {Lim}, {Lupton}, {Marshall}, {Marshall}, {May},
  {Miller}, {Mills}, {Monet}, {Neill}, {Nordby}, {O'Connor}, {Oliver},
  {Olivier}, {Olsen}, {Owen}, {Peterson}, {Petry}, {Pierfederici},
  {Pietrowicz}, {Pike}, {Pinto}, {Plante}, {Radeka}, {Rasmussen}, {Ridgway},
  {Rosing}, {Saha}, {Schalk}, {Schindler}, {Schneider}, {Schumacher}, {Sebag},
  {Seppala}, {Shipsey}, {Silvestri}, {Smith}, {Smith}, {Strauss}, {Stubbs},
  {Sweeney}, {Szalay}, {Thaler}, {Vanden Berk}, {Walkowicz}, {Warner},
  {Willman}, {Wittman}, {Wolff}, {Wood-Vasey}, {Yoachim}, {Zhan}, \& {for the
  LSST Collaboration}}]{Ivezic2008a}
{Ivezic}, Z., {Tyson}, J.~A., {Acosta}, E., {et~al.} 2008, ArXiv e-prints,
  arXiv:0805.2366

\bibitem[{{Jones}(1981)}]{Jones1981a}
{Jones}, R.~H. 1981, in Applied Time Series Analysis III, ed. D.~{Findley} (New
  York, NY: Academic Press)

\bibitem[{Jones \& Ackerson(1990)}]{Jones1990a}
Jones, R.~H., \& Ackerson, L.~M. 1990, Biometrika, 77, pp. 721

\bibitem[{{Kaiser} {et~al.}(2002){Kaiser}, {Aussel}, {Burke}, {Boesgaard},
  {Chambers}, {Chun}, {Heasley}, {Hodapp}, {Hunt}, {Jedicke}, {Jewitt},
  {Kudritzki}, {Luppino}, {Maberry}, {Magnier}, {Monet}, {Onaka}, {Pickles},
  {Rhoads}, {Simon}, {Szalay}, {Szapudi}, {Tholen}, {Tonry}, {Waterson}, \&
  {Wick}}]{Kaiser2002a}
{Kaiser}, N., {Aussel}, H., {Burke}, B.~E., {et~al.} 2002, in Society of
  Photo-Optical Instrumentation Engineers (SPIE) Conference Series, Vol. 4836,
  Survey and Other Telescope Technologies and Discoveries, ed. J.~A. {Tyson} \&
  S.~{Wolff}, 154--164

\bibitem[{{Kelly} {et~al.}(2009){Kelly}, {Bechtold}, \&
  {Siemiginowska}}]{Kelly2009a}
{Kelly}, B.~C., {Bechtold}, J., \& {Siemiginowska}, A. 2009, \apj, 698, 895

\bibitem[{{Kelly} {et~al.}(2011){Kelly}, {Sobolewska}, \&
  {Siemiginowska}}]{Kelly2011a}
{Kelly}, B.~C., {Sobolewska}, M., \& {Siemiginowska}, A. 2011, \apj, 730, 52

\bibitem[{{Kelly} {et~al.}(2013){Kelly}, {Treu}, {Malkan}, {Pancoast}, \&
  {Woo}}]{Kelly2013b}
{Kelly}, B.~C., {Treu}, T., {Malkan}, M., {Pancoast}, A., \& {Woo}, J.-H. 2013,
  \apj, 779, 187

\bibitem[{{Kochanek}(2004)}]{Kochanek2004a}
{Kochanek}, C.~S. 2004, \apj, 605, 58

\bibitem[{{Koen}(2005)}]{Koen2005a}
{Koen}, C. 2005, \mnras, 361, 887

\bibitem[{{Koz{\l}owski} {et~al.}(2010){Koz{\l}owski}, {Kochanek}, {Udalski},
  {Wyrzykowski}, {Soszy{\'n}ski}, {Szyma{\'n}ski}, {Kubiak}, {Pietrzy{\'n}ski},
  {Szewczyk}, {Ulaczyk}, {Poleski}, \& {OGLE Collaboration}}]{Kozowski2010a}
{Koz{\l}owski}, S., {Kochanek}, C.~S., {Udalski}, A., {et~al.} 2010, \apj, 708,
  927

\bibitem[{{Law} {et~al.}(2009){Law}, {Kulkarni}, {Dekany}, {Ofek}, {Quimby},
  {Nugent}, {Surace}, {Grillmair}, {Bloom}, {Kasliwal}, {Bildsten}, {Brown},
  {Cenko}, {Ciardi}, {Croner}, {Djorgovski}, {van Eyken}, {Filippenko}, {Fox},
  {Gal-Yam}, {Hale}, {Hamam}, {Helou}, {Henning}, {Howell}, {Jacobsen},
  {Laher}, {Mattingly}, {McKenna}, {Pickles}, {Poznanski}, {Rahmer}, {Rau},
  {Rosing}, {Shara}, {Smith}, {Starr}, {Sullivan}, {Velur}, {Walters}, \&
  {Zolkower}}]{Law2009a}
{Law}, N.~M., {Kulkarni}, S.~R., {Dekany}, R.~G., {et~al.} 2009, \pasp, 121,
  1395

\bibitem[{{Liu}(2004)}]{Liu2004a}
{Liu}, J. 2004, Monte Carlo Strategies in Scientific Computing (New York, NY:
  Springer)

\bibitem[{{MacLeod} {et~al.}(2010){MacLeod}, {Ivezi{\'c}}, {Kochanek},
  {Koz{\l}owski}, {Kelly}, {Bullock}, {Kimball}, {Sesar}, {Westman}, {Brooks},
  {Gibson}, {Becker}, \& {de Vries}}]{MacLeod2010a}
{MacLeod}, C.~L., {Ivezi{\'c}}, {\v Z}., {Kochanek}, C.~S., {et~al.} 2010,
  \apj, 721, 1014

\bibitem[{{MacLeod} {et~al.}(2011){MacLeod}, {Brooks}, {Ivezi{\'c}},
  {Kochanek}, {Gibson}, {Meisner}, {Koz{\l}owski}, {Sesar}, {Becker}, \& {de
  Vries}}]{MacLeod2011a}
{MacLeod}, C.~L., {Brooks}, K., {Ivezi{\'c}}, {\v Z}., {et~al.} 2011, \apj,
  728, 26

\bibitem[{{MacLeod} {et~al.}(2012){MacLeod}, {Ivezi{\'c}}, {Sesar}, {de Vries},
  {Kochanek}, {Kelly}, {Becker}, {Lupton}, {Hall}, {Richards}, {Anderson}, \&
  {Schneider}}]{MacLeod2012a}
{MacLeod}, C.~L., {Ivezi{\'c}}, {\v Z}., {Sesar}, B., {et~al.} 2012, \apj, 753,
  106

\bibitem[{Marquardt \& Stelzer(2007)}]{Marquardt2007a}
Marquardt, T., \& Stelzer, R. 2007, Stochastic Processes and their
  Applications, 117, 96

\bibitem[{{Miller} {et~al.}(2010){Miller}, {Turner}, {Reeves}, {Lobban},
  {Kraemer}, \& {Crenshaw}}]{Miller2010a}
{Miller}, L., {Turner}, T.~J., {Reeves}, J.~N., {et~al.} 2010, \mnras, 403, 196

\bibitem[{{Morgan} {et~al.}(2012){Morgan}, {Hainline}, {Chen}, {Tewes},
  {Kochanek}, {Dai}, {Kozlowski}, {Blackburne}, {Mosquera}, {Chartas},
  {Courbin}, \& {Meylan}}]{Morgan2012a}
{Morgan}, C.~W., {Hainline}, L.~J., {Chen}, B., {et~al.} 2012, \apj, 756, 52

\bibitem[{{Mushotzky} {et~al.}(2011){Mushotzky}, {Edelson}, {Baumgartner}, \&
  {Gandhi}}]{Mushotzky2011a}
{Mushotzky}, R.~F., {Edelson}, R., {Baumgartner}, W., \& {Gandhi}, P. 2011,
  \apjl, 743, L12

\bibitem[{{Nowak}(2000)}]{Nowak2000a}
{Nowak}, M.~A. 2000, \mnras, 318, 361

\bibitem[{{Pancoast} {et~al.}(2011){Pancoast}, {Brewer}, \&
  {Treu}}]{Pancoast2011a}
{Pancoast}, A., {Brewer}, B.~J., \& {Treu}, T. 2011, \apj, 730, 139

\bibitem[{{Pancoast} {et~al.}(2012){Pancoast}, {Brewer}, {Treu}, {Barth},
  {Bennert}, {Canalizo}, {Filippenko}, {Gates}, {Greene}, {Li}, {Malkan},
  {Sand}, {Stern}, {Woo}, {Assef}, {Bae}, {Buehler}, {Cenko}, {Clubb},
  {Cooper}, {Diamond-Stanic}, {Hiner}, {H{\"o}nig}, {Joner}, {Kandrashoff},
  {Laney}, {Lazarova}, {Nierenberg}, {Park}, {Silverman}, {Son}, {Sonnenfeld},
  {Thorman}, {Tollerud}, {Walsh}, \& {Walters}}]{Pancoast2012a}
{Pancoast}, A., {Brewer}, B.~J., {Treu}, T., {et~al.} 2012, \apj, 754, 49

\bibitem[{{Press} {et~al.}(1992){Press}, {Rybicki}, \& {Hewitt}}]{Press1992a}
{Press}, W.~H., {Rybicki}, G.~B., \& {Hewitt}, J.~N. 1992, \apj, 385, 404

\bibitem[{{Reimann}(1994)}]{Reimann1994a}
{Reimann}, J.~D. 1994, PhD thesis, UNIVERSITY OF CALIFORNIA, BERKELEY.

\bibitem[{Roux(2002)}]{roux2002some}
Roux, A. 2002, PhD thesis, University of Pretoria

\bibitem[{{Ruan} {et~al.}(2012){Ruan}, {Anderson}, {MacLeod}, {Becker},
  {Burnett}, {Davenport}, {Ivezi{\'c}}, {Kochanek}, {Plotkin}, {Sesar}, \&
  {Stuart}}]{Ruan2012a}
{Ruan}, J.~J., {Anderson}, S.~F., {MacLeod}, C.~L., {et~al.} 2012, \apj, 760,
  51

\bibitem[{{Rybicki} \& {Press}(1992)}]{Rybicki1992a}
{Rybicki}, G.~B., \& {Press}, W.~H. 1992, \apj, 398, 169

\bibitem[{{Rybicki} \& {Press}(1995)}]{Rybicki1995a}
---. 1995, Physical Review Letters, 74, 1060

\bibitem[{Sanderson(2010)}]{Sanderson2010a}
Sanderson, C. 2010, Technical Report, NICTA, 2

\bibitem[{{Scargle}(1982)}]{Scargle1982a}
{Scargle}, J.~D. 1982, \apj, 263, 835

\bibitem[{Schlemm \& Stelzer(2012)}]{Schlemm2012a}
Schlemm, E., \& Stelzer, R. 2012, Bernoulli, 18, 46

\bibitem[{{Sesar} {et~al.}(2010){Sesar}, {Ivezi{\'c}}, {Grammer}, {Morgan},
  {Becker}, {Juri{\'c}}, {De Lee}, {Annis}, {Beers}, {Fan}, {Lupton}, {Gunn},
  {Knapp}, {Jiang}, {Jester}, {Johnston}, \& {Lampeitl}}]{Sesar2010a}
{Sesar}, B., {Ivezi{\'c}}, {\v Z}., {Grammer}, S.~H., {et~al.} 2010, \apj, 708,
  717

\bibitem[{{Soszy{\'n}ski} {et~al.}(2011){Soszy{\'n}ski}, {Udalski},
  {Szyma{\'n}ski}, {Kubiak}, {Pietrzy{\'n}ski}, {Wyrzykowski}, {Ulaczyk},
  {Poleski}, {Koz{\l}owski}, \& {Pietrukowicz}}]{Soszynski2011a}
{Soszy{\'n}ski}, I., {Udalski}, A., {Szyma{\'n}ski}, M.~K., {et~al.} 2011, Acta
  Astronomica, 61, 217

\bibitem[{Spiegelhalter {et~al.}(2002)Spiegelhalter, Best, Carlin, \& Van
  Der~Linde}]{spiegelhalter2002bayesian}
Spiegelhalter, D.~J., Best, N.~G., Carlin, B.~P., \& Van Der~Linde, A. 2002,
  Journal of the Royal Statistical Society: Series B (Statistical Methodology),
  64, 583

\bibitem[{{Uttley} {et~al.}(2002){Uttley}, {McHardy}, \&
  {Papadakis}}]{Uttley2002a}
{Uttley}, P., {McHardy}, I.~M., \& {Papadakis}, I.~E. 2002, \mnras, 332, 231

\bibitem[{{Uttley} {et~al.}(2005){Uttley}, {McHardy}, \&
  {Vaughan}}]{Uttley2005a}
{Uttley}, P., {McHardy}, I.~M., \& {Vaughan}, S. 2005, \mnras, 359, 345

\bibitem[{{Vaughan}(2010)}]{Vaughan2010a}
{Vaughan}, S. 2010, \mnras, 402, 307

\bibitem[{{Vaughan} {et~al.}(2003){Vaughan}, {Edelson}, {Warwick}, \&
  {Uttley}}]{Vaughan2003a}
{Vaughan}, S., {Edelson}, R., {Warwick}, R.~S., \& {Uttley}, P. 2003, \mnras,
  345, 1271

\bibitem[{Vihola(2012)}]{Vihola2012a}
Vihola, M. 2012, Statistics and Computing, 22, 997

\bibitem[{{Vio} {et~al.}(1992){Vio}, {Cristiani}, {Lessi}, \&
  {Provenzale}}]{Vio1992a}
{Vio}, R., {Cristiani}, S., {Lessi}, O., \& {Provenzale}, A. 1992, \apj, 391,
  518

\bibitem[{Wang(2013)}]{wang2013cts}
Wang, Z. 2013, Journal of Statistical Software, 53, 1

\bibitem[{{Zu} {et~al.}(2013){Zu}, {Kochanek}, {Koz{\l}owski}, \&
  {Udalski}}]{Zu2013a}
{Zu}, Y., {Kochanek}, C.~S., {Koz{\l}owski}, S., \& {Udalski}, A. 2013, \apj,
  765, 106

\bibitem[{{Zu} {et~al.}(2011){Zu}, {Kochanek}, \& {Peterson}}]{Zu2011a}
{Zu}, Y., {Kochanek}, C.~S., \& {Peterson}, B.~M. 2011, \apj, 735, 80

\end{thebibliography}

\appendix

\section{Kalman Filter for a CARMA Process}
\label{s-kfilter}

Using the rotated state space representation, the Kalman Filter computes the mean and variance of the measured time series at time $t_i$ conditional on the measurements at times $\{ t_j ; j<i \}$ via the following algorithm \citep{Jones1990a}:
\begin{enumerate}
\item
Center the time series. For each $i$ compute $\tilde{y}_i = y_i - \mu$. Because the Kalman Filter assumes a zero-mean time series, we will work with the centered values instead of $y_i$.
\item
Denote the covariance matrix of the predicted rotated state as $\tilde{P}$. Initialize the rotated state vector $\tilde{\bf x}$ and its covariance $\tilde{P}$ at time $t_1$ to its stationary mean and covariance, ${\bf 0}$ and $\tilde{V}$ respectively:
\begin{align}
	\tilde{\bf x}(t_1|\cdot)& = {\bf 0} \\
	\tilde{P}(t_1|\cdot)& = \tilde{V}
\end{align}
Defining ${\bf J} = U^{-1} {\bf e}$, the stationary covariance matrix for $\tilde{\bf x}(t)$ has elements \citep{Belcher1994a}
\begin{equation}
	\tilde{V}_{lk} = -\frac{J_l J^*_j}{r_l + r^*_k} \label{eq-rot_stationary_covar}.
\end{equation}

\item
Calculate the mean and variance of the first measurement in the time series using the stationary values for a CARMA process:
\begin{align}
	E(\tilde{y}_1)& = 0 \\
	Var(\tilde{y}_1|\sigma,\alpha,\beta)& = R(0) + \sigma^2_1,
\end{align}
where $R(0)$ is given by Equation (\ref{eq-carma_autocovar}).

\item
Initialize the Kalman gain:
\begin{equation}
	{\bf K}_1 = \frac{\tilde{P}(t_1|\cdot) \tilde{\bf b}^H}{Var(\tilde{y}_1|\sigma,\alpha,\beta)}.
\end{equation}
Here ${\bf z}^H$ denotes the Hermitian transpose of ${\bf z}$.

\item
Update the estimate of the rotated state vector:
\begin{equation}
	\tilde{\bf x}(t_1|t_1) = \tilde{\bf x}(t_1|\cdot) + \tilde{y}_1 {\bf K}_1.
\end{equation}

\item
Update the covariance matrix of the rotated state vector:
\begin{equation}
	\tilde{P}(t_1|t_1) = \tilde{P}(t_1|\cdot) - Var(\tilde{y}_1|\sigma,\alpha,\beta) {\bf K}_1 {\bf K}_1^H.
\end{equation}

\end{enumerate}

After the initializing the Kalman filter as above, repeat the following steps for $i = 2, \ldots, n$:
\begin{enumerate}[resume]
\item 
Predict the rotated state vector at the next observation time given the time series at the earlier observation times:
\begin{equation}
	\tilde{\bf x}(t_i|t_{i-1}) = \Lambda_i \tilde{\bf x}(t_{i-1}|t_{i-1}).
	\label{eq-state_transition}
\end{equation}

\item
Calculate the covariance matrix of the predicted rotated state vector at time $t_i$:
\begin{equation}
	\tilde{P}(t_i|t_{i-1}) = \Lambda_i (\tilde{P}(t_{i-1}|t_{i-1}) - \tilde{V}) \Lambda_i^H + \tilde{V}
	\label{eq-rot-state_covar}
\end{equation}

\item
Calculate the mean and variance of the centered time series at time $t_i$ conditional on the earlier values:
\begin{align}
	E(\tilde{y}_i|\tilde{\bf y}_{<i}, \sigma, \alpha, \beta)& = \tilde{\bf b} \tilde{\bf x}(t_i|t_{i-1}) \label{eq-ycmean} \\
	Var(\tilde{y}_i | \tilde{\bf y}_{<i}, \sigma, \alpha, \beta)& = \tilde{\bf b} \tilde{P}(t_i|t_{i-1}) \tilde{\bf b}^H + \sigma^2_i.
	\label{eq-ycvar}
\end{align}
Here we have used the notation $\tilde{\bf y}_{<i} = [\tilde{y}_1, \ldots, \tilde{y}_{i-1}]$.

\item
Update the Kalman gain:
\begin{equation}
	{\bf K}_i = \frac{\tilde{P}(t_i|t_{i-1}) \tilde{\bf b}^H}{Var(\tilde{y}_i|\tilde{\bf y}_{<i}, \sigma,\alpha,\beta)}.
	\label{eq-kalman_gain}
\end{equation}

\item
Update the estimated rotated state vector:
\begin{equation}
	\tilde{\bf x}(t_i|t_i) = \tilde{\bf x}(t_i|t_{i-1}) + (\tilde{y}_i - E(\tilde{y}_i|\tilde{\bf y}_{<i}, \sigma, \alpha, \beta)) {\bf K}_i.
	\label{eq-updated_rot_state}
\end{equation}

\item
Finally, update the covariance matrix of the estimated rotated state vector:
\begin{equation}
\tilde{P}(t_i|t_i) = \tilde{P}(t_i|t_{i-1}) - Var(\tilde{y}_i|\tilde{\bf y}_{<i},\sigma,\alpha,\beta) {\bf K}_i {\bf K}_i^H.
\label{eq-updated_covar}
\end{equation}

\end{enumerate}
The values of $E(\tilde{y}_i | \tilde{\bf y}_{<i}, \sigma, \alpha, \beta)$ and $Var(\tilde{y}_i | \tilde{\bf y}_{<i},\sigma,\alpha,\beta)$ computed using the above algorithm can then be used to efficiently calculate the likelihood function given by Equation (\ref{eq-likhood}), noting that $E(y_i|{\bf y}_{<i}, \sigma, \alpha, \beta, \mu) = E(\tilde{y}_i | \tilde{\bf y}_{<i}, \sigma, \alpha, \beta) + \mu$.
	
\section{Algorithm for Computing the Coefficients Needed for Interpolation and Extrapolation from a Measured Time Series Under the CARMA Model}
\label{s-coefs}

The coefficients $\tilde{c}_i$ and $\tilde{d}_i$ needed to compute the expected value of $y_i$ as a function of $y_0$ for $i \geq j(t_0)$ can be computed recursively using the following algorithm:
\begin{enumerate}
\item
First, run the Kalman filter up to index $j(t_0)-1$. If $t_0 < t_1$ then skip this step.

\item
Compute $\tilde{\bf x}(t_0|t_{j(t_0)-1}), \tilde{P}(t_0|t_{j(t_0)-1}), E(\tilde{y}_0|\tilde{\bf y}_{<j(t_0)},\theta),$ and $Var(\tilde{y}_0|\tilde{\bf y}_{<j(t_0)}, \theta)$ using Equations (\ref{eq-state_transition})--(\ref{eq-ycvar}). If $t_0 > t_n$, then nothing further needs to be calculated. Otherwise, use these values to compute ${\bf K}_0, \tilde{\bf x}(t_0|t_0)$, and $\tilde{P}(t_0|t_0)$.

\item
Initialize the rotated state vector coefficients ${\bf c}_{j(t_0)}$ and ${\bf d}_{j(t_0)}$ as
\begin{align}
	{\bf c}_{j(t_0)} & = \Lambda_0 [\tilde{\bf x}(t_0|t_0) - E(\tilde{y}_0|\tilde{\bf y}_{<j(t_0)},\theta) {\bf K}_0] \\
	{\bf d}_{j(t_0)} & = \Lambda_0 {\bf K}_0,
\end{align}
where $\Lambda_0$ is a diagonal matrix with $\Lambda_{0,kk} = e^{r_k (t_{j(t_0)} - t_0)}$. If $t_0 < t_1$ then $j(t_0) = 1$ and the initial values are
\begin{align}
	{\bf c}_1 & = {\bf 0} \\
	{\bf d}_1 & = \frac{\tilde{V} \tilde{\bf b}^H}{\tilde{\bf b} \tilde{V} \tilde{\bf b}^H}
\end{align}

\item
Initialize the coefficients $\tilde{c}_{j(t_0)}$ and $\tilde{d}_{j(t_0)}$ as
\begin{align}
	\tilde{c}_{j(t_0)} & = \tilde{\bf b} {\bf c}_{j(t_0)} \\
	\tilde{d}_{j(t_0)} & = \tilde{\bf b} {\bf d}_{j(t_0)}
\end{align}
Note that if $t_0 < t_1$ then $\tilde{c}_1 = 0$ and $\tilde{d}_1 = 1$.
\end{enumerate}

Then, for $i = j(t_0)+1, \ldots, n$ do
\begin{enumerate}[resume]
\item
Update the linear coefficients for the rotated state vector
\begin{align}
	{\bf c}_{i} & = \Lambda_i [ {\bf c}_{i-1} + (y_{i-1} - \tilde{c}_{i-1}) {\bf K}_i ] \\
	{\bf d}_{i} & = \Lambda_i [ {\bf d}_{i-1} - \tilde{d}_{i-1} {\bf K}_i ].
\end{align}

\item
Update the linear coefficients:
\begin{align}
	\tilde{c}_i & = \tilde{\bf b} {\bf c}_i \\
	\tilde{d}_i & = \tilde{\bf b} {\bf d}_i
\end{align}
\end{enumerate}
Because the Kalman gains, ${\bf K}_i$, only depend on the observation times, and not on the measured time series, they are computed by performing the Kalman filter using the observation times $t_1, \ldots, t_{j(t_0)-1}, t_0, t_{j(t_0)}, \ldots, t_n$.

\end{document}